\documentclass[11pt]{article}
\usepackage{amsmath}
\usepackage{amssymb}
\usepackage{epsfig}
\newlength{\bredde}
\def\slash#1{\settowidth{\bredde}{$#1$}\ifmmode\,\raisebox{.15ex}{/}
\hspace*{-\bredde} #1\else$\,\raisebox{.15ex}{/}\hspace*{-\bredde} #1$\fi}
\newcommand{\be}{\begin{equation}}
\newcommand{\ee}{\end{equation}}
\newcommand{\bea}{\begin{eqnarray}}
\newcommand{\eea}{\end{eqnarray}}
\newcommand{\nn}{\nonumber}

\newcommand{\tP}{\tilde{\varphi}}
\newcommand{\Lh}{\hat{L}^{\nu}}
\newcommand{\Lhe}{\hat{L}^{1}}
\newcommand{\Lhb}{\hat{L}^{\bar{\nu}}}

\newcommand{\one}{\mbox{\bf 1}}

\def\Sl#1{\rlap{\raisebox{.15ex}{$\mskip 4 mu /$}}#1}  

\newcommand{\sect}[1]{\setcounter{equation}{0}\section{#1}}

\def\hx{\hat{x}}
\def\hy{\hat{y}}
\def\hm{\hat{m}}
\def\hs{\hat{s}}
\def\hmu{\hat{\mu}}
\def\hd{\hat{\delta}}

\begin{document}
\topmargin -1.4cm
\oddsidemargin -0.8cm
\evensidemargin -0.8cm
\title{\Large{{\bf
The $k$th Smallest Dirac Operator Eigenvalue \\and the Pion Decay Constant
}}}

\vspace{1.5cm}
\author{~\\{\sc G.~Akemann}$^1$ and {\sc A.~C.~Ipsen}$^{2}$
\\~\\$^1$Department of Physics,
Bielefeld University,
Postfach 100131,
D-33501 Bielefeld, Germany
\\~\\
$^2$Niels Bohr International Academy and Discovery Center,
Niels Bohr Institute,\\
Blegdamsvej 17, DK-2100 Copenhagen {\O}, Denmark}

\date{}

\maketitle
\vfill
\begin{abstract}
We derive an analytical expression for the distribution of the $k$th
smallest Dirac eigenvalue in QCD with imaginary isospin chemical
potential in the Dirac operator
for arbitrary gauge field topology $\nu$.
Because of its dependence on
the pion decay constant $F_{\pi}$ through the chemical potential
in the epsilon-regime of chiral perturbation theory
this can be used for lattice determinations of that low-energy constant.
On the technical side we use a chiral Random-Two Matrix Theory, where
we express the $k$th eigenvalue distribution
through the joint probability of the ordered $k$ smallest
eigenvalues. The latter can be computed exactly for finite and
infinite $N$, for which we derive generalisations of Dyson's
integration Theorem and Sonine's identity.

\end{abstract}
\vfill

\thispagestyle{empty}
\newpage

\renewcommand{\thefootnote}{\arabic{footnote}}
\setcounter{footnote}{0}

\sect{Introduction}\label{intro}

It is by now well known how the spontaneous breaking of chiral
symmetry in QCD leads to remarkably strong predictions for the spectral
properties of the Dirac operator in that theory. Based first
exclusively on the relation to the effective field theory for
the associated Nambu-Goldstone bosons at fixed gauge field topology
\cite{LS}, an intriguing relation to universal Random Matrix Theory
(RMT) was
also pointed out \cite{Jacetal}. It has subsequently become clear how
these two alternative formulations are related, and all $n$-point
spectral correlation functions have been shown to be identical
in these two formulations to leading order in a $1/L$-expansion
(where $L \equiv V^{1/4}$ gives the extent of the space-time volume
$V$) \cite{DOTV,Basile}.
This holds then also for individual distributions of Dirac
operator eigenvalues \cite{AD}.

If one seeks sensitivity to the pion decay constant $F_{\pi}$
it turns out to be useful to consider the Dirac operator of
quark doublets with isospin chemical potential $\mu$.
Based on the chiral Lagrangian formulation \cite{Imiso}, it has been
suggested to use a spectral 2-point function of the two associated
Dirac operators with imaginary isospin chemical potential. The
advantage of imaginary chemical potential lies in the fact that
the corresponding Dirac operator retains its anti-hermiticity.
To leading order, all results can be expressed in terms of the
simple finite-volume scaling variable $\hat{\mu} = \mu F_{\pi}\sqrt{V}$.
In this way, $F_{\pi}$ can be extracted from fits that vary $\mu$
and/or $V$. There is also sensitivity to $F_{\pi}$ in other observables
that couple to chemical potential \cite{Kim,MT,Luz,Tilo,Laurent}.

The leading-order chiral Lagrangian computations of ref.
\cite{Imiso} have been given a reformulation in terms of a chiral
Random Two-Matrix Theory in ref. \cite{ADOS}. In this way,
all spectral correlation functions
associated with the two Dirac operators
${\cal D}_1$ and ${\cal D}_2$, with respective chemical potentials $\mu_1$ and $\mu_2$,
have been computed analytically
in \cite{ADOS} for both the quenched and the full theory with
$N_f$ light flavours. It also includes
all spectral correlation functions where the imaginary isospin
chemical potential only enters in the Dirac operator whose eigenvalues
are being computed,
while the gauge field configurations are obtained in
the usual way at vanishing chemical potential. In analogy with what
is being done when varying quark masses away from the value used
for generating the gauge field configurations we call this
``partial quenching''.

In an earlier paper \cite{AD08} it was shown how all probability
distributions of individual Dirac operator eigenvalues can be computed
by means of a series expansion in higher $n$-point spectral correlation
functions.
In reference \cite{AD08} an
explicit analytical formula was also given for the lowest non-zero eigenvalue
distribution for arbitrary combinations
of flavours $N_1$ and $N_2$ of the Dirac operators ${\cal D}_1$ and ${\cal D}_2$, respectively,
at gauge
field topology $\nu=0$, in an approach quite close
to that of ref. \cite{DNW}. Two obvious questions remained open that
we will answer in this paper: how to extend this to non-zero topology
$\nu>0$, and how to compute the distribution of the second, third or
general $k$th eigenvalue as these are known for zero chemical
potential \cite{DN}.
However, the path of ref. \cite{AD08} is not very
suitable in particular
for the derivation of the distributions of higher eigenvalues
in a compact analytical manner.
Our  setup will follow closely ref. \cite{DN} at
vanishing chemical potential. We shall present here a new formalism
that immediately allows for the analytical determination of
the distributions of these higher eigenvalues.
The extension of the first approach \cite{AD08} to $\nu>0$ for the
first eigenvalue is
presented in appendix \ref{oldnugen}, as an alternative formulation
and analytical check to part of our new approach.
The benefit of our new results should be two-fold: while expressions for higher topology $\nu>0$ allow for an independent determination of $\Sigma$ and $F_\pi$ from different lattice configurations, the expressions for higher eigenvalues should allow for a better determination using the same configurations as for the first eigenvalue.

How much of the program proposed in \cite{Imiso} to determine $F_{\pi}$ on
the lattice has been realised in the meantime?
Based on a preliminary account \cite{AD07} of ref. \cite{AD08}, the
expansion of the first eigenvalue was first used in simulations in
\cite{Tom}.
However, the question remained how large the finite-volume corrections
to the leading order (LO) $\epsilon$-expansion are, in which the chiral
Lagrangian-RMT correspondence holds. In a series of papers
this question has been addressed and answered:
in \cite{CT1} the next to LO corrections (NLO) and in \cite{CT2}
next to next to LO (NNLO) corrections in the epsilon-expansion were
computed.
As a result of these computations at NLO all RMT expressions for arbitrary $n$-point density correlations functions
(and thus for all individual eigenvalues too) remain valid. The infinite volume expressions simply get renormalised by finite-volume corrections, one only has to
replace $\Sigma$ and $F_{\pi}$ by $\Sigma_{ef\!f}$ and $F_{\pi\,ef\!f}$ in the corresponding RMT expressions. Here
the subscript ``$ef\!f$'' for effective encodes the corrections
that match those computed earlier in \cite{GL} and
\cite{DGH,Laurent}, respectively.
Only at NNLO non-universal, non-RMT
corrections appear. It was further noticed in \cite{CT1,CT2}, that the
size of the corrections at each order depends considerably on the lattice
geometry, in particular when using asymmetric geometries.
In order to keep the NNLO corrections small,
in \cite{CT3} the authors used a specific optimised
geometry where they could apply RMT predictions
and effective couplings at NLO only, and they obtained realistic values for $\Sigma$ and $F_{\pi}$ from partially quenched lattice data
for small chemical potential.
Technically speaking
$\Sigma$ was determined there from the first eigenvalue distribution at vanishing chemical potential, and $F_{\pi}$ from the shift of the eigenvalues compared to zero chemical potential. We refer to \cite{CT3} for a more detailed discussion of these fits.

Motivated by these findings we have completed the computation for the
$k$th Dirac eigenvalues for all $\nu\geq0$ in the RMT setting, in order
to have a more complete mathematical toolbox at hand.

Our paper is organised as follows. In the next section we briefly
define the notation and remind the reader of the definition of
chiral Random Two-Matrix Theory. We introduce a certain
joint probability density and describe how it can be used to
derive individual eigenvalue distributions. We give the explicit
finite-$N$ solution here in terms of new polynomials and a new
sequence of matrix model kernels. We take the scaling limit relevant for
QCD in section \ref{scaling},
and write out explicitly and discuss the physically most important
examples such as partially quenched $N_f=2$ results.
Section \ref{conc} contains our conclusions and a suggestion for a quite
non-trivial but important extension of these results.
Because some of the relevant technical details have been described in
ref. \cite{AD08}, we have relegated many of the mathematical
details in this
paper to appendices. In addition, in Appendix A we describe
an explicit construction of the polynomials needed to compute
the first eigenvalue distribution in sectors of non-trivial gauge
field topology if one alternatively uses the method of ref. \cite{AD08}.

\sect{Chiral Random Two-Matrix Theory}\label{defs}

Before turning to the relevant Random Two-Matrix Theory, we first
briefly outline the set-up in the language of the gauge field theory.
We are considering QCD at finite four-volume $V$, and we assume that
chiral symmetry is spontaneously broken at infinite volume. We consider
two Dirac operators $D_{1,2}$ with different imaginary
baryon (quark) chemical potential
$\mu_{1,2}$,
\bea
D_1\psi_1^{(n)} &\equiv & [\Sl{D}(A)+i\mu_1\gamma_0]\psi_1^{(n)} ~=~
 i\lambda_1^{(n)}\psi_1^{(n)}  \cr
D_2\psi_2^{(n)} &\equiv & [\Sl{D}(A)+i\mu_2\gamma_0]\psi_2^{(n)} ~=~
 i\lambda_2^{(n)}\psi_2^{(n)} ~.
\label{Ddef}
\eea
When $\mu \equiv \mu_1 = -\mu_2$ this is simply imaginary isospin
chemical potential, but we can stay with the more general case.
We thus consider $N_1$ light quarks
coupled to quark chemical potential $\mu_1$, and $N_2$ light quarks
coupled to quark chemical potential $\mu_2$. Let us first consider
the conceptually simplest case where $N_1+N_2=N_f$, and later comment
on the changes needed to deal with partial quenching.

In the chiral Lagrangian framework the terms that depend on $\mu_{1,2}$
are easily written down on the basis of the usual correspondence
with external vector sources. Going to the $\epsilon$-regime of
chiral perturbation theory in sectors of fixed gauge field
topology \cite{LS}, the leading term in the effective partition
function including imaginary $\mu_{1,2}$ reads \cite{Imiso,Son} (see also \cite{Kogut} for QCD-like theories)
\be
{Z}_{\nu}^{(N_f)} =
\int_{U(N_f)} dU \,(\det U)^{\nu}e^{\frac{1}{4}VF_\pi^2
{\rm Tr} [U,B][U^\dagger,B] + \frac{1}{2} \Sigma V
{\rm Tr}({\cal M}^\dagger U + {\cal M}U^\dagger)} ~.
\label{ZXPT}
\ee
In (\ref{ZXPT}) the $N_f\times N_f$ matrix
\be
B ~=~ {\rm diag}(\mu_1\one_{N_1},\mu_2\one_{N_2})
\ee
is made out of the chemical potentials, and
the quark mass matrix is
\be
{\cal M} ~=~ \mbox{diag}(m_1,\ldots,m_{N_f}) ~.
\ee
The partition function (\ref{ZXPT}) is a simple zero-dimensional
group integral.
The leading contribution to the
effective low-energy field theory at finite volume $V$ in the
$\epsilon$-regime is thus well known.

We consider now the limit in which $V \to \infty$ while $\hat{m} = m\Sigma V$
and $\hat{\mu} = \mu F_{\pi} \sqrt{V}$ are kept
fixed. In this limit, to LO in the $\epsilon$-expansion,
the effective partition function of this
theory and all the spectral correlation functions of its Dirac
operator eigenvalues are completely
equivalent to  the chiral Random Two-Matrix Theory with imaginary chemical
potential that was introduced in ref. \cite{ADOS}. The equivalence for the
two-point function follows from \cite{Imiso}, for all higher density
correlations it was proven in \cite{Basile}. Therefore, since we have
proven \cite{AD08} that the probability distribution of the $k$th
smallest eigenvalue can be computed in terms of this infinite sequence
of spectral correlation functions, we are free to use the chiral
Random Two-Matrix Theory when performing the actual analytical computation.

As already mentioned it has been shown in \cite{CT1} that also to NLO in the
$\epsilon$-expansion the random matrix expressions \cite{ADOS} for density correlation
functions remain valid, when
replacing $\Sigma$ and $F_{\pi}$ by the renormalised constants $\Sigma_{ef\!f}$ and $F_{\pi\,ef\!f}$
that encode the finite-volume corrections.
Only to NNLO non-universal corrections to
the random matrix setting appear.

The partition
function of chiral
Random Two-Matrix Theory is, up to an irrelevant normalisation factor,
defined as
\bea
{\cal Z}_{\nu}^{(N_f)} &=&
 \int d\Phi  d\Psi~
e^{-{N}{\rm Tr}\left(\Phi^{\dagger}
\Phi + \Psi^{\dagger}\Psi\right)}
\prod_{f_1=1}^{N_1} \det[{\cal D}_1 + m_{f_1}]
\prod_{f_2=1}^{N_2} \det[{\cal D}_2 + m_{f_2}]
\label{ZNf}
\eea
where ${\cal D}_{1,2}$ are given by
\bea
{\mathcal D}_{1,2} = \left( \begin{array}{cc}
0 & i \Phi + i \mu_{1,2} \Psi \\
i \Phi^{\dagger} + i \mu_{1,2} \Psi^{\dagger} & 0
\end{array} \right) ~.
\eea
The operator remains
anti-Hermitian because the chemical potentials are imaginary, as shown
explicitly.
Both $\Phi$ and $\Psi$ are complex rectangular
matrices of size $N\times (N+\nu)$, where both $N$ and $\nu$ are integers.
The index $\nu$ corresponds to gauge field topology in the usual way.
The aforementioned correspondence to chiral perturbation theory holds in the following
microscopic large-$N$ limit:
\be
\lim_{N\to\infty}{\cal Z}_{\nu}^{(N_f)}\ =\ {Z}_{\nu}^{(N_f)} \, \ \
\mbox{with}\ \ \hat{m} = 2Nm\ ,\ \
\hat{\mu} = \sqrt{2N}\ \mu \ .
\ee

In the framework of chiral Random Two-Matrix Theory it is particularly
simple to consider the situation corresponding to what we
call partial quenching. Here one simply
considers eigenvalues of one of the matrices, say  $D_{1}$, that then does
not enter into the actual integration measure of (\ref{ZNf}) by
setting $N_1=0$.
In the language
of the chiral Lagrangian, this needs to be done in terms of
graded groups or by means of the replica method.

Referring to ref. \cite{ADOS} for details, we immediately write down
the corresponding representation in terms of eigenvalues
$x_i^2$ and $y_i^2$ of
${\cal D}_1$ and ${\cal D}_2$, respectively,
\bea
{\cal Z}_{\nu}^{(N_f)} &=& \int_0^{\infty} \prod_{i=1}^Ndx_idy_i\
{\cal P}_{\nu}^{(N_f)}(\{x\},\{y\}) \ ,
\label{evrep}
\eea
up to an irrelevant (mass dependent) normalization factor. The integrand is
the {\em joint probability distribution function} (jpdf),
which is central for what follows:
\bea
{\cal P}_{\nu}^{(N_f)}(\{x\},\{y\})
&\equiv& \prod_{i=1}^N\left((x_iy_i)^{\nu+1}e^{-N(c_1 x_i^2 + c_2 y_i^2)}
\prod_{f_1=1}^{N_1} (x_i^2+m_{f_1}^2)
\prod_{f_2=1}^{N_2} (y_i^2+m_{f_2}^2) \right) \cr
&&\cr
&&\times\ \Delta_N(\{x^2\})\Delta_N(\{y^2\})\det_{1\leq i,j\leq
  N}\left[I_{\nu}(2 d N x_i y_j)\right]  \ .
\label{jpdf}
\eea

Because the integration in eq. (\ref{ZNf})
was over $\Phi$ and $\Psi$ separately, the matrices now
become coupled in the exponent. The corresponding unitary group integral
leads to the determinant of modified $I$-Bessel functions, and removes
one of the initially two  Vandermonde determinants, which is  defined as
$\Delta_N(\{x^2\})=\prod_{j>i=1}^N(x_j^2-x_i^2)$.
The precise connection between the constants and $\mu_{1,2}$ is given by
\bea
c_1 &=& (1+\mu_2^2)/\delta^2  \ ,\ \ \ \ c_2 \ =\ (1+\mu_1^2)/\delta^2 \ ,\nn\\
d &=& (1+\mu_1\mu_2)/\delta^2 \ ,\ \ \delta \ =\ \mu_2 - \mu_1 \ ,\nn\\
1-\tau &=& d^2/(c_1c_2)\ ,
\label{cdtdef}
\eea
where the latter is defined for later convenience. We need the
joint probability distribution to be normalised to unity, which
is done trivially by dividing by ${\cal Z}_{\nu}^{(N_f)}$ (cf. eq.
\ref{evrep})).

\sect{The $k$th Eigenvalue at Finite-$N$ for Arbitrary
  $\nu\geq0$}\label{sectkth}

We now follow the derivation of ref. \cite{DN} rather closely. We are
able to do that because we focus here on the distributions of individual
$x$-eigenvalues only - which are those we may partially quench.
For that purpose it is convenient to first consider
the joint probability distribution of the $k$ smallest $x$-eigenvalues,
ordered such that $0 \leq x_1 \leq x_2 \leq \ldots \leq x_k$:
\bea
\Omega_{\nu}^{(N_f)}(x_1, \ldots, x_k)
& \equiv & \frac{N!}{{\cal Z}_{\nu}^{(N_f)}(N-k)!}
\int_{x_k}^{\infty}\!dx_{k+1} \cdots
\int_{x_k}^{\infty}\!dx_N \int_0^{\infty}\prod_{i=1}^N dy_i\
{\cal P}_{\nu}^{(N_f)}(\{x\},\{y\}).
\label{kjpdf}
\eea
This quantity is then used to generate the $k$th $x$-eigenvalue distribution
through the following integration\footnote{Compared to \cite{DN} we are
already working with squared variables here. Translating to that
picture the integration bounds in eqs. (\ref{kjpdf}) and (\ref{pkdef})
remain the same as in  \cite{DN}.}
\bea
p_{k}^{(N_f,\,\nu)}(x_k) = \int_0^{x_k} dx_1 \int_{x_1}^{x_k} dx_2\ldots
 \int_{x_{k-2}}^{x_k} dx_{k-1}\
\Omega_{\nu}^{(N_f)}(x_1, \ldots, x_k)  \ .
\label{pkdef}
\eea
Note that for $k=1$ no integration is needed, and
$p_{1}^{(N_f,\,\nu)}(x_1)= \Omega_{\nu}^{(N_f)}(x_1)$.

The computation of mixed or conditional individual eigenvalue distributions,
e.g. to find the joint distribution of the first $x$- and first $y$-eigenvalue,
remains an open problem.

We next proceed as in ref. \cite{AD08}, and integrate out {\it all}
$y$-eigenvalues
exactly. Because of this we note that in eq. (\ref{kjpdf})
we can replace the determinant over the Bessel functions by
$N!$ times its diagonal part, after having made use of the antisymmetry
property of $\Delta_N(y^2)$.
After inserting a representation of the Bessel function in terms of
a factorised infinite sum over Laguerre polynomials (see eq. (B.7) in
\cite{ADOS}), we get
\bea
&&\int_0^{\infty}\prod_{i=1}^N dy_i\
{\cal  P}_{\nu}^{(N_f)}(\{x\},\{y\}) =
N!\int_0^{\infty}\prod_{i=1}^N \left(dy_i
\prod_{f_1=1}^{N_1} (x_i^2+m_{f_1}^2)\prod_{f_2=1}^{N_2}(y_i^2+m_{f_2}^2)\right)
\Delta_N(\{x^2\}) \nn\\
&&\times \Delta_N(\{y^2\})
\prod_{i=1}^N\left(\!(N d)^\nu \tau^{\nu+1}
(x_i y_i)^{2\nu+1} e^{-N\tau (c_1 x_i^2 + c_2 y_i^2)}
 \sum_{n_i=0}^{\infty} \frac{n_i!(1-\tau)^{n_i}}{(n_i+\nu)!}
 L_{n_i}^{\nu} (N \tau c_1 x_i^2) L_{n_i}^{\nu} (N \tau c_2 y_i^2)
\!\!\right)\!,\nn\\
&&\label{intP}
\eea
where the Laguerre polynomials $L_j^\nu(N\tau c_2y^2)$
now appear with their corresponding weight function
$y^{2\nu+1} e^{-N\tau c_2 y^2}$
due to the identity used.
Next we include the set of $N_2$ masses, $\{m_2\}$, into  $\Delta_N(\{y^2\})$
to form a larger Vandermonde determinant of size $N+N_2$, and then replace
it by a determinant of in general arbitrary Laguerre polynomials
normalised to be monic
\be
\Delta_N(\{y^2\})\prod_{i=1}^N\prod_{f_2=1}^{N_2}(y_i^2+m_{f_2}^2)
=
\frac{\left|
\begin{array}{ccc}
\Lhb_0(N\tau c_2(im_{f_2=1})^2)& \cdots
&\frac{1}{(N\tau  c_2)^{N+N_2-1}}\Lhb_{N+N_2-1}(N\tau c_2(im_{f_2=1})^2)\\
\cdots&\cdots &\cdots\\
\Lhb_0(N\tau c_2(im_{N_2})^2)& \cdots
&\frac{1}{(N\tau c_2)^{N+N_2-1}}\Lhb_{N+N_2-1}(N\tau c_2(im_{N_2})^2)\\
\Lhb_0(N\tau c_2y_1^2)&\cdots
&\frac{1}{(N\tau c_2)^{N+N_2-1}}\Lhb_{N+N_2-1}(N\tau c_2y_1^2)\\
\cdots&\cdots &\cdots\\
\Lhb_0(N\tau c_2y_N^2)&\cdots
&\frac{1}{(N\tau c_2)^{N+N_2-1}}\Lhb_{N+N_2-1}(N\tau c_2y_N^2)\\
\end{array}
\!\right|}{\Delta_{N_2}(\{(im_2)^2\})}.
\label{Delta}
\ee
Here the index of the Laguerre polynomials
$\bar{\nu}$ is arbitrary.
The monic Laguerre polynomials relate to
ordinary Laguerre polynomials in a $\nu$-independent manner:
\be
\Lh_n(x)\equiv (-1)^n n!\ L_n^\nu(x)\ =\ \sum_{j=0}^n (-1)^{n+j}
\frac{n!(n+\nu)!}{(n-j)!(\nu+j)!j!} \ x^j\ =\ x^n+{\cal O}(x^{n-1})\ \ .
\label{Lmonic}
\ee
In eq. (\ref{Delta}) the inverse powers $(N\tau c_2)^{j-1}$ can be taken out
of the determinant.
Inserting this back into eq. (\ref{intP}) for $\nu=\bar{\nu}$
we can use the orthogonality of the
Laguerre polynomials in the integrated variables $y_{1,\ldots,N}$,
killing the infinite sums from the expanded Bessel functions. The Laguerre
polynomials in $x$ thus replace those in $y$ inside the determinant, times
the norm from the integration. We obtain
\bea
&&\int_0^{\infty}\prod_{i=1}^N dy_i\ {\cal P}_{\nu}^{(N_f)}(\{x\},\{y\})
\ = \ \cr
&=&\frac{N!(N d)^{N\nu} \tau^{N(\nu+1)}
\prod_{j=0}^{N+N_2-1} (1-\tau)^j(N\tau c_2)^{-j}}
{\Delta_{N_2}(\{(im_2)^2\})\ 2^N(N\tau c_2)^{N(\nu+1)}}
\prod_{i=1}^N \left( x_i^{2\nu+1} e^{-N\tau c_1 x_i^2}
\prod_{f_1=1}^{N_1} (x_i^2+m_{f_1}^2)\right)
\Delta_N(\{x^2\})\cr
&&\times
\left|
\begin{array}{ccc}
\Lh_0(N\tau c_2(im_{f_2=1})^2)& \cdots &
\frac{1}{(1-\tau)^{N+N_2-1}}\Lh_{N+N_2-1}(N\tau c_2(im_{f_2=1})^2)
\\
\cdots & \cdots &\cdots\\
\Lh_0(N\tau c_2(im_{N_2})^2)& \cdots &
\frac{1}{(1-\tau)^{N+N_2-1}}\Lh_{N+N_2-1}(N\tau c_2(im_{N_2})^2)
\\
\Lh_0(N\tau c_1x_1^2)& \cdots &
\Lh_{N+N_2-1} (N \tau c_1 x_1^2)
\\
\cdots & \cdots &\cdots\\
\Lh_0(N\tau c_1x_N^2)& \cdots &
\Lh_{N+N_2-1} (N \tau c_1 x_N^2)
\\
\end{array}
\right|,
\label{intP2}
\eea
after taking out common factors of the determinant.
The determinant in eq. (\ref{intP2}), which we call $D_{N+N_2}$,
can almost be mapped to a Vandermonde determinant, using an identity proved in
appendix A in \cite{AD08}
\bea
D_{N+N_2}(\{m_2^2\};\{x^2\})&=&\left|
\begin{array}{ccc}
\Lh_0(\frac{1}{\tau}{M_{f_2=1}^2})& \cdots &
\frac{\tau^{N+N_2-1}}{(1-\tau)^{N+N_2-1}} \Lh_{N+N_2-1}
(\frac{1}{\tau}{M_{f_2=1}^2})
\\
\cdots & \cdots &\cdots\\
\Lh_0(\frac{1}{\tau}{M_{N_2}^2})& \cdots &
\frac{\tau^{N+N_2-1}}{(1-\tau)^{N+N_2-1}} \Lh_{N+N_2-1}
(\frac{1}{\tau}{M_{N_2}^2})\\
1& \cdots &
X_1^{2(N+N_2-1)}
\\
\cdots & \cdots &\cdots\\
1
& \cdots &
X_N^{2(N+N_2-1)}
\\
\end{array}
\right|,\cr
&&
\label{Ldelta}
\eea
where we have defined
\be
M_{f_2}^2\equiv N \tau c_2(im_{f_2})^2\ \ \mbox{and} \ \
X_j^2\equiv N \tau c_1 x_j^2\ .
\ee
This fact can be used below
to perform the $N-k$ remaining integrations in the generating
quantity $\Omega_k$, after inserting eq. (\ref{Ldelta})
into eqs. (\ref{intP2}) and (\ref{kjpdf}). This leads to
\bea
\Omega_{\nu}^{(N_f)}(x_1, \ldots, x_k)  &=&
C\prod_{j>i\geq1}^{k} (x_j^2-x_i^2)\
\int_{x_k}^{\infty}\!dx_{k+1} \cdots
\int_{x_k}^{\infty}\!dx_N
\prod_{j>i\geq k+1}^{N} (x_j^2-x_i^2)
\prod_{j=k+1}^N\prod_{i=1}^k(x_j^2-x_i^2)
\cr
&&\times
\prod_{i=1}^N \left( x_i^{2\nu+1}  e^{-N\tau c_1 x_i^2}
\prod_{f_1=1}^{N_1}(x_i^2+m_{f_1}^2)\right)
D_{N+N_2}(\{m_2^2\};\{x^2\})\ ,
\label{Omdelta}
\eea
where we have split the Vandermonde determinant
$\Delta_N(\{x^2\})$ into integrated and
unintegrated variables, and defined the following constant
\be
C\ \equiv\ \frac{(N!)^2(N d)^{N\nu} \tau^{N(\nu+1)}
\prod_{j=0}^{N+N_2-1}(1-\tau)^j(N\tau c_2)^{-j}}
{{\cal Z}_\nu^{(N_f)}(N-k)!\ 2^N(N\tau c_2)^{N(\nu+1)}
\Delta_{N_2}(\{(im_2)^2\})}\ .
\label{Cdef}
\ee
We can now change variables $x_j\to u_j=x_j^2$ for $j=k+1,\ldots,N$,
and then perform the shift
$u_j\to z_j=u_j-x_k^2$ to obtain integrations $\int_0^\infty dz_j$ in
eq. (\ref{Omdelta}):
\bea
\Omega_{\nu}^{(N_f)}(x_1, \ldots, x_k)  &=&
C\prod_{j>i\geq1}^{k}\! (x_j^2-x_i^2) \prod_{i=1}^k\!
\left( x_i^{2\nu+1}  e^{-N\tau c_1 x_i^2}
\prod_{f_1=1}^{N_1}(x_i^2+m_{f_1}^2)\right)\frac{1}{2^{(N-k)}}
e^{-N(N-k)\tau c_1x_k^2}
\cr
&\times&
\int_{0}^{\infty}\prod_{j=k+1}^N\left( dz_j\ z_j\,  e^{-N\tau c_1z_j}\
(z_j+x_k^2)^{\nu}\prod_{i=1}^{k-1}(z_j+x_k^2-x_i^2)
\prod_{f_1=1}^{N_1}(z_j+x_k^2+m_{f_1}^2)\right)
\cr
&&\cr
&\times& \prod_{j>i\geq k+1}^{N}(z_j-z_i)\
D_{N+N_2}(\{m_2^2\};x_1^2,\ldots,x_k^2,z_{k+1}+x_k^2,\ldots,z_N+x_k^2 )\ .
\label{Omdelta2}
\eea
We thus obtain an integral with $\nu+k-1$ extra mass terms of flavour-type
``1'', in addition to the $N_1$ shifted masses.
The weight
\be
w(z)=z^1 e^{-N\tau c_1z} \label{L1weight}
\ee
is now of Laguerre-type corresponding to a
{\em fixed} topological charge of $\bar{\nu}=1$, irrespective of the actual
topological charge $\nu$ of the given gauge field sector we started with. We will
therefore call $\bar{\nu}$ spurious topology.
Compared with
the corresponding derivation in case of vanishing chemical potential
\cite{DN}, this can be seen to differ by one unit,
compared to spurious topology $\bar{\nu}=2$ at vanishing $\mu_{1,2}$ in \cite{DN}. The
reason for this difference is easily traced to the different
integration measure for the $x$-eigenvalues, which has one power
less in the Vandermonde determinant compared to the case of
vanishing chemical potential\footnote{Of course, the additional pieces due
to the $y$-integrations are what ensures equivalence to those corresponding
one-matrix model results in the limit $\mu_j\to0$.}.
It is an interesting and quite non-trivial
check on our present calculation that we recover the results of
reference \cite{DN} in the limit of vanishing chemical potential. In particular,
the shift from spurious topology $\bar{\nu}=1$ to spurious topological
charge $\bar{\nu}=2$ in the integration measure
will now arise due to recurrence relations of Laguerre
polynomials. Some details of this will be given below.

When replacing the Vandermonde determinant in the variables $z_j$ as well as
$D_{N+N_2}$ by a determinant containing Laguerre polynomials we thus
choose polynomials
$L_j^1(N\tau c_1 z)$ in order to be able to exploit the orthogonality
properties with respect to the measure $w(z)$ eq. (\ref{L1weight}).

For the new masses times $\Delta_{N-k}(\{z\})$ this is an easy task. We can
include them into a bigger determinant of size $N-k+N_1+\nu+k-1$, following
the identity eq. (\ref{Delta}). Here we replace the $N$ variables $y_i^2$ by
$N-k$ variables $z_i$, and the set of $N_2$ masses by the following set of
$N_1+\nu+k-1$ masses:
\bea
m_{f_1}^{\prime\,2} &\equiv m_{f_1}^2+x_k^2&\ \ \mbox{for}\ \ {f_1=1,\ldots, N_1}\
, \cr
m_{N_1+j}^{\prime\,2} &\equiv\ \ x_k^2+\epsilon_j^2&\ \ \mbox{for}
\ \ {j=1,\ldots, \nu}\ ,
\cr
m_{N_1+\nu+i}^{\prime\,2} &\equiv\ \  x_k^2-x_i^2&\ \ \mbox{for}\ \ {i=1,\ldots,
  k-1}\ ,
\label{defmf1shift}
\eea
and likewise we define
\be
M_{j}^{\prime\,2}\equiv N \tau c_1(im_{j}^\prime)^2\ \ \mbox{for} \ \
j=1,\ldots,N_1+\nu+k-1\ .
\ee
For computational simplicity we first set the $\nu$ degenerate masses to be
different by adding small pairwise different constants, $\epsilon_j^2$,
and then set
$\epsilon_j=0$ at the end of the computation.
Also we may chose spurious topology $\bar{\nu}=1$ in eq. (\ref{Delta}).
The prefactors in front of the Laguerre polynomials inside the determinant can
be taken out.

To express the determinant $D_{N+N_2}$ of the shifted arguments in
eq. (\ref{Omdelta2}) in terms of Laguerre polynomials
requires a bit more algebra:
\bea
&&D_{N+N_2}(\{m_2^2\};x_1^2,\ldots,x_k^2,z_{k+1}+x_k^2,\ldots,z_N+x_k^2 )\ =\cr
&&\cr
&&=
\left|
\begin{array}{ccc}
\Lh_0(\frac{1}{\tau}M_{f_2=1}^2)&\cdots &
\sum_{l=0}^{N+N_2-1}\frac{\tau^l}{(1-\tau)^{l}}\Lh_l (\frac{1}{\tau}M_{f_2=1}^2)
(-X_k^2)^{N+N_2-1-l} {N+N_2-1 \choose l}
\\
\cdots & \cdots &\cdots\\
\Lh_0(\frac{1}{\tau}M_{N_2}^2)&\cdots &
\sum_{l=0}^{N+N_2-1}\frac{\tau^l}{(1-\tau)^{l}} \Lh_l (\frac{1}{\tau}M_{N_2}^2)
(-X_k^2)^{N+N_2-1-l} {N+N_2-1 \choose l}
\\
1& \cdots & (X_1^2-X_{k}^2)^{N+N_2-1}
\\
\cdots & \cdots &\cdots\\
1& \cdots & (X_{k-1}^2-X_{k}^2)^{N+N_2-1}
\\
1& \cdots & 0\\
1& \cdots & Z_{k+1}^{N+N_2-1}
\\
\cdots & \cdots &\cdots\\
1& \cdots & Z_N^{N+N_2-1}
\\
\end{array}
\right|
\cr
&&\nn\\
&&
=\left|
\begin{array}{ccc}
q_0^\nu(M_{f_2=1}^2) & \cdots &q_{N+N_2-1}^\nu(M_{f_2=1}^2)
\\
\cdots &\cdots &\cdots\\
q_0^\nu(M_{N_2}^2) & \cdots &q_{N+N_2-1}^\nu(M_{N_2}^2)
\\
\hat{L}_0^1(M_{N_1+\nu+1}^{\prime\,2}) &\cdots& \hat{L}_{N+N_2-1}^1(M_{N_1+\nu+1}^{\prime\,2})
\\
\cdots &\cdots &\cdots\\
\hat{L}_0^1(M_{N_1+\nu+k}^{\prime\,2}) &\cdots& \hat{L}_{N+N_2-1}^1(M_{N_1+\nu+k}^{\prime\,2})
\\
\hat{L}_0^1(Z_{k+1}) & \cdots & \hat{L}_{N+N_2-1}^1(Z_{k+1})
\\
\cdots &\cdots &\cdots\\
\hat{L}_0^1(Z_N) & \cdots & \hat{L}_{N+N_2-1}^1(Z_N)
\\
\end{array}
\right|\ ,
\label{Lfinal}
\eea
where for convenience we have defined $M_{N_1+\nu+k}^{\prime\,2} = 0$,
as well as
\be
Z_k\equiv N\tau c_1 z_k \ .
\ee
In the first step in eq. (\ref{Lfinal})
we have used the invariance of the determinant to undo the
shift in $x_k^2$ of the $z_j$ variables. This leads to a shift in variables
$X_{i}$ and to linear combinations of the Laguerre polynomials in the $N_2$
masses. In the second step we have added columns from the left to the right to
replace monic powers in $Z_j$ and $M_i^{\prime\,2}$ by polynomials
$\hat{L}_n^1$. Because the determinant $D_{N+N_2}$
is not an invariant Vandermonde
this leads to a further sum in the first $N_2$ rows, invoking the following
{\em new polynomials}
\be
q_n^\nu(M_2^2)\ =\ (-)^n n!\sum_{l=0}^n\frac{1}{(1-\tau)^l}
L_l^\nu(M_2^2)L_{n-l}^{-\nu}(-X_k^2)\ .
\label{qfinal}
\ee
The form given here is derived in Appendix \ref{Qid}
using identities for Laguerre polynomials. For later purpose,
we note already that in the limit of zero
chemical potential, $i.e.$ in the limit $\tau\to0$, we obtain Laguerre
polynomials of shifted mass from the $q_n^\nu$:
\be
\lim_{\tau\to0}q_n^\nu(M_2^2)\frac{1}{(-)^n n!}\ =\
\sum_{j=0}^n L_j^\nu(-Nm_2^2)L_{n-j}^{-\nu}(-Nx_k^2)
\ =\ L_n^1\left(-N(m_2^2+x_k^2)\right) \ .
\label{Lsum}
\ee
In this way we recover, after the use of a few identities for Laguerre
polynomials, the results of ref. \cite{DN} in the limit
of vanishing chemical potential.

We now proceed with the integration over the variables $z_{k+1},\ldots,z_N$ in
eq. (\ref{Omdelta2}). Using the rewriting discussed above, we have:
\bea
&&\!\!\!\!\!\!\!\!\Omega_{\nu}^{(N_f)}(x_1, \ldots, x_k) =
C\prod_{j>i\geq1}^{k}\! (x_j^2-x_i^2) \prod_{i=1}^k\!
\left( x_i^{2\nu+1}  e^{-N\tau c_1 x_i^2}
\prod_{f_1=1}^{N_1}(x_i^2+m_{f_1}^2)\right)
\frac{2^{-(N-k)}e^{-N(N-k)\tau c_1x_k^2}}{
\Delta_{N_1+\nu+k-1}(\{(im^\prime)^2\})}
\cr
&&\!\!\!\!\times
\int_{0}^{\infty}\prod_{j=k+1}^N\left( dz_j\ z_j\,  e^{-N\tau c_1z_j}\right)
\prod_{j=0}^{N+N_1+\nu-2} (N\tau c_1)^{-j}
\cr
&&\!\!\!\!\times
\left|\!
\begin{array}{ccc}
\Lhe_0(M_1^{\prime\,2})&\cdots &\Lhe_{N+N_1+\nu-2}(M_1^{\prime\,2})\\
\cdots&\cdots &\cdots\\
\Lhe_0(M_{N_1+\nu+k-1}^{\prime\,2})&\cdots
&\Lhe_{N+N_1+\nu-2}(M_{N_1+\nu+k-1}^{\prime\,2})\\
\Lhe_0(Z_{k+1})&\cdots &\Lhe_{N+N_1+\nu-2}(Z_{k+1})\\
\cdots&\cdots &\cdots\\
\Lhe_0(Z_{N})&\cdots &\Lhe_{N+N_1+\nu-2}(Z_{N})\\
\end{array}
\!\right|
\left|\!
\begin{array}{ccc}
q_0^\nu(M_{f_2=1}^2) & \cdots &q_{N+N_2-1}^\nu(M_{f_2=1}^2)
\\
\cdots &\cdots &\cdots\\
q_0^\nu(M_{N_2}^2) & \cdots &q_{N+N_2-1}^\nu(M_{N_2}^2)
\\
\hat{L}_0^1(M_{N_1+\nu+1}^{\prime\,2}) &\cdots& \hat{L}_{N+N_2-1}^1(M_{N_1+\nu+1}^{\prime\,2})
\\
\cdots &\cdots &\cdots\\
\hat{L}_0^1(M_{N_1+\nu+k}^{\prime\,2}) &\cdots& \hat{L}_{N+N_2-1}^1(M_{N_1+\nu+k}^{\prime\,2})
\\
\hat{L}_0^1(Z_{k+1}) & \cdots & \hat{L}_{N+N_2-1}^1(Z_{k+1})
\\
\cdots &\cdots &\cdots\\
\hat{L}_0^1(Z_N) & \cdots & \hat{L}_{N+N_2-1}^1(Z_N)
\\
\end{array}
\!\right|\!.\cr
&&\label{Omdelta3}
\eea
This expression can be simplified somewhat by noting the identity
\begin{multline}
  \frac{
  \prod_{j>i\geq1}^{k}\! (x_j^2-x_i^2)
  \prod_{i=1}^k\!
    \left(x_i^{2\nu+1}
    \prod_{f_1=1}^{N_1}(x_i^2+m_{f_1}^2)\right)}
  {\Delta_{N_1+\nu+k-1}(\{(im^\prime)^2\})}\\
  = \frac{
  x_k^{2\nu+1}\prod_{i=1}^{k-1}\bigl(x_i[x_k^2-x_i^2]\bigr)
  \prod_{f_1=1}^{N_1}(x_k^2+m_{f_1}^2)}
  {\Delta_{N_1}(\{(im_1)^2\})\Delta_{\nu}(\{(i\epsilon)^2\})
  \prod_{f_1=1}^{N_1}m_{f_1}^{2\nu}}
  \bigl(1+O(\epsilon)\bigr)\ .
\label{Omsimp}
\end{multline}
The Vandermonde determinant
$\Delta_{N_1}(\{(im_1)^2\})$ also occurs as a factor in the
partition function, and thus cancels.

The integrals over $z_j, j = k+1, \ldots, N$ can be performed, again
exploiting orthogonality properties of the Laguerre polynomials $L_n^1(z)$
with respect to our weight function (\ref{L1weight}). This is done
by means of a generalisation of the original Dyson Theorem (see
e.g. \cite{Mehta}), now for two determinants of different size, of
which different sets of entries are not even
of Laguerre-type, and not integrated over either.
The needed generalisation of this theorem
was essentially provided in ref. \cite{AV}, and we only need a slight
extension of this more general theorem here. We have relegated
the proof of the new theorem to Appendix \ref{AVgen}, and will only
quote the result here.

To express the result, we define two kernels (both corresponding to
the kernel $K^{I}$ of Appendix \ref{AVgen} for different values of the indices)
\newcommand{\kerL}{K}
\newcommand{\kerq}{{\tilde K}}
\newcommand{\kermat}{\mathcal K}
\begin{align}
  \kerL(M_1^{\prime\, 2},M_2^{\prime\, 2}) &\equiv \sum_{j=0}^\Lambda
  \frac 1 {j+1} L_j^1(M_1^{\prime\, 2})L_j^1(M_2^{\prime\, 2})\ , &
  \kerq^\nu(M^{\prime\, 2},M^2) &\equiv \sum_{j=0}^\Lambda
  \frac{(-)^j}{(j+1)!} L_j^1(M^{\prime\, 2})q_j^\nu(M^2)\ ,
\label{Knudef}
\end{align}
with $\Lambda = \min(N+N_1+\nu-2,N+N_2-1)$, and the associated matrix
of
size $(N_2+k)\times(N_1+\nu+k-1)$
\be
  \kermat \equiv \begin{pmatrix}
    \kerq^\nu(M_1^{\prime\,2},M_{f_2=1}^2)&\cdots&\kerq^\nu(M_{N_1+\nu+k-1}^{\prime\,2},M_{f_2=1}^2)\\
    \cdots&\cdots&\cdots\\
    \kerq^\nu(M_1^{\prime\,2},M_{N_2}^2)&\cdots&\kerq^\nu(M_{N_1+\nu+k-1}^{\prime\,2},M_{N2}^2)\\
    \kerL(M_1^{\prime\,2},M_{N_1+\nu+1}^{\prime\,2})&\cdots&\kerL(M_{N_1+\nu+k-1}^{\prime\,2},M_{N_1+\nu+1}^{\prime\,2})\\
    \cdots&\cdots&\cdots\\
    \kerL(M_1^{\prime\,2},M_{N_1+\nu+k}^{\prime\,2})&\cdots&\kerL(M_{N_1+\nu+k-1}^{\prime\,2},M_{N_1+\nu+k}^{\prime\,2})
  \end{pmatrix}\ .
\ee

We can now give the answer for $\Omega_{\nu}^{(N_f)}$, where we have
to distinguish 3 different cases, depending on the respective sizes of
the two determinants in \eqref{Omdelta3}.
\begin{itemize}

\item[$i)$]
The simplest case is $N_2 = N_1+\nu-1$ (this implies that $\kermat$ is square).
Applying Appendix \ref{AVgen} to the integral of \eqref{Omdelta3} and
using eq. \eqref{Omsimp} we obtain to leading order in mass difference $\epsilon_j$:
\begin{multline}
\Omega_{\nu}^{(N_f)}(x_1, \ldots, x_k) =
C \frac{
  x_k^{2\nu+1}\prod_{i=1}^{k-1}\bigl(x_ie^{-N\tau c_1 x_i^2}[x_k^2-x_i^2]\bigr)
  \prod_{f_1=1}^{N_1}(x_k^2+m_{f_1}^2)}
  {2^{N-k}\Delta_{N_1}(\{(im_1)^2\})\Delta_{\nu}(\{(i\epsilon)^2\})
  \prod_{f_1=1}^{N_1}m_{f_1}^{2\nu}}
 (N\tau c_1)^{-2(N-k)} \label{intfinal} \\
 \times e^{-N(N-k+1)\tau c_1x_k^2}\prod_{j=0}^{N+N_1+\nu-2} (N\tau c_1)^{-j}
  \prod_{j=0}^\Lambda
  \biggl((j+1)!j!\biggr)(N-k)!(-)^{(N_1+N_2+\nu+1)(N-k)}\det \kermat
\end{multline}
A typical example for this case would be $N_1=N_2=\nu=1$.

\item[$ii)$]For $N_2>N_1+\nu-1$ the expression is the same, but with additional
columns in the determinant in the second line of eq. (\ref{intfinal}):
\be
  \left|\begin{array}{cc}
    \kermat &
    \begin{array}{ccc}
      q_{N+N_1+\nu-1}^\nu(M_{f_2}^2)&\cdots &q_{N+N_2-1}^\nu(M_{f_2}^2)\\
      \Lhe_{N+N_1+\nu-1}(M_{N_1+\nu+j}^{\prime\,2})&\cdots&
\Lhe_{N+N_2-1}(M_{N_1+\nu+j}^{\prime\,2})
    \end{array}
  \end{array}\right|\ ,\quad f_2 = 1,\ldots,N_2,\  j=1,\ldots,k\ .
  \label{intfinalbigN2}
\ee
This case typically occurs for partial quenching $N_1=0$, $N_2=2$ and
small topology $\nu=0,1,2$

\item[$iii)$]
Similarly, for $N_2<N_1+\nu-1$, which
typically occurs for higher topology, the determinant
in eq. (\ref{intfinal})
is replaced by
\be
  \left|\begin{array}{cc}
    \kermat^T&
    \begin{array}{ccc}
      \Lhe_{N+N_2}(M_j^{\prime\,2})&\cdots&\Lhe_{N+N_1+\nu-2}(M_j^{\prime\,2})
    \end{array}
  \end{array}\right|\ ,\quad j=1,\ldots,N_1+\nu+k-1\ .
  \label{intfinalsmallN2}
\ee
\end{itemize}
We note that all three cases contain degeneracies for $\nu>1$, and that a
Taylor expansion must be performed when sending $\epsilon_j\to0$.

Finally we are ready to compute the $k$th individual eigenvalue
probability distribution $p_k^{(\nu)}(x)$ by means of inserting the
above results into eq. (\ref{pkdef})
\bea
p_k^{(N_f,\,\nu)}(x_k) ~=~ \int_0^{x_k}\! dx_1\int_{x_1}^{x_k}\!dx_2 \cdots \int_{x_{k-2}}^{x_k}
dx_{k-1}
\Omega_{\nu}^{(N_f)}(x_1, \ldots, x_k) ~.
\nn
\eea
An alternative, equivalent formulation for the first eigenvalue
$p_{k=1}^{(\nu)}(x)$ at arbitrary $\nu$ follows from the derivative of
its cumulative
distribution given in appendix \ref{oldnugen}, where we generalise the
approach of \cite{AD08} to $\nu>0$.

We refrain from giving further explicit examples for finite $N$
and instead turn to the large-$N$ limit.

\sect{The Microscopic Scaling Limit}\label{scaling}

Having obtained the explicit solution for any finite $N$, we are now
ready to take the appropriate microscopic $N \to \infty$ scaling limit.
In the language of QCD this corresponds to a finite-volume scaling
in $V$, where $V$ is the space-time volume. It is only in this limit that
we expect to obtain universal results which do not depend on having chosen
Gaussian measures for the two original matrices $\Phi$ and $\Psi$.
We remind the reader that the appearance of Laguerre polynomials
in the finite-$N$ solution is directly linked to these two original
measures having been chosen Gaussian. As we have seen,
there are numerous instances where our derivation makes
explicit reference to specific identities for Laguerre polynomials,
and also the measure factor was specific to these Gaussian integrals.
It is therefore a quite non-trivial task, and an interesting
challenge, to generalise the universality proof of the last reference
in \cite{Jacetal} to this more general setting. The universality has
been implicitly checked by the equivalence proof of \cite{Basile},
but an explicit proof of universality directly
in the framework of the chiral Random Two-Matrix Theory remains to be
constructed.

In QCD-terminology, we keep $\hat{m}_i \equiv m_i\Sigma V$ and
$\hat{\mu} \equiv \mu F_{\pi}\sqrt{V}$ fixed, while we take $V \to \infty$.
In the language of our chiral Random Two-Matrix Theory, we take
the $N \to \infty$ limit while keeping
\bea
\hat{x}_i \equiv 2Nx_i ~~~,~~~~~~~
\hm_i \equiv 2Nm_i ~~~,~~~~~~~ \hat{\mu}_{1,2}^2 \equiv
2N\mu_{1,2}^2
\eea
fixed (we also scale the masses $\epsilon_j$ in the same way, before taking them
to zero at the end of the calculation). In addition, we introduce the following relevant quantity, the
{\em difference} in rescaled chemical potential,
$\hat{\delta} \equiv \hat{\mu}_2 - \hat{\mu}_1$. We will follow
\cite{AD08} closely, without giving a detailed derivation.
Defining the following continuum indices
\be
  t\equiv j/N\text{, and } r\equiv l/j\ ,
\ee
we replace the sum by an integral, $\sum_{j=0}^\Lambda N^{-1}\to\int_0^1 dt$, and
correspondingly for index $l$. In
addition, we use the identities
\be
  \lim_{N,j,l\to\infty} \frac 1 {(1-\tau)^l} = \exp\left[\frac 1 2 rt\hat\delta^2\right]\ ,
  \quad
  \lim_{N,j,l\to\infty} j^{-\nu}L_l^\nu(M^{\prime\,2}=-N\tau c_1m^2)
    = \left(\frac {4r}{\hat m^2 t}\right)^{\nu/2} I_\nu(\sqrt{rt}\hat m)\ .
\ee
The scaling limit of the kernel $\kerL$ is obtained using the Christoffel-Darboux
identity:
\begin{align}
  \kerL_S(\hm'_1,\hm'_2) &\equiv  \lim_{N\to\infty}\frac 1 {N^2}
    \kerL_N(M_1^{\prime\,2},M_2^{\prime\,2})\nonumber\\
  &= \lim_{N\to\infty}-\frac 1 {N^2}
    \frac{L_{N+1}^1(M_1^{\prime\,2})L_N^1(M_2^{\prime\,2})
      -L_{N+1}^1(M_2^{\prime\,2})L_N^1(M_1^{\prime\,2})}
     {M_1^{\prime\,2}-M_2^{\prime\,2}}\nonumber\\
  &= \frac 8 {\hm'_1\hm'_2}\frac{\hm'_1I_0(\hm'_1)I_1(\hm'_2)-\hm'_2I_0(\hm'_2)I_1(\hm'_1)}
    {\hm_1^{\prime\,2}-\hm_2^{\prime\,2}}
\end{align}

The scaling limit of the new polynomials $q^\nu_N$ requires a little more
care. For $\nu>0$ one needs to treat the terms corresponding
to indices $l = N-\nu+1,\ldots,N$ in \eqref{qfinal}
separately, due to the presence
of Laguerre polynomials with negative index\footnote{For integer $\nu>0$ and
  fixed $j,z$, the limit $\lim_{n\to\infty} n^\nu L_j^{-\nu}(z/n)$ is only finite for
  $j\geq\nu$.} \cite{AD08}.
For these ``anomalous'' terms, the scaling limit is
\begin{align}
  q_{SA}^\nu(\hm;t) &\equiv \lim_{N,j\to\infty} j^{-1}
    \sum_{l=j-\nu+1}^{j}\frac{1}{(1-\tau)^l}L_l^\nu(M^2)L_{j-l}^{-\nu}(-X_k^2)\nonumber\\
  &= \lim_{N,j\to\infty}\sum_{p=0}^{\nu-1}\frac{j^{\nu-p-1}}{p!}
    \left(-\frac{t\hx_k^2}{4}+O(j^{-1})\right)^p
    \sum_{l=j-\nu+1}^{j-p}(-)^{j-l}{\nu-p-1 \choose j-l-p}
    \frac{1}{(1-\tau)^l}j^{-\nu}L_l^\nu(M^2)\nonumber\\
  &=\sum_{p=0}^{\nu-1}\frac 1 {p!}\left(\frac{t\hx_k^2}{4}\right)^p
    \left.\frac{\partial^{\nu-p-1}}{\partial r^{\nu-p-1}}\right|_{r=1}
    \left[e^{\frac 1 2r t\hd^2}\left(\frac{4r}{t\hm^2}\right)^{\nu/2}I_\nu(\sqrt{rt}\hm)\right]
    \ .
  \label{qSA}
\end{align}
Here we write out $L^{-\nu}_{j-l}$, and use the fact that the binomial
weight kills the terms of order lower than $\nu-p-1$ of the expansion
in $r$. Note that naively $q_{SA}^\nu$ is proportional to $j^{\nu-1}$, but that the
pattern of cancelation is exactly such that the limit is finite.
The notation in eq. (\ref{qSA}) is chosen such that for $\nu=0$ the
sum is void, and thus the anomalous term is absent for $\nu=0$, $q_{SA}^{\nu=0}=0$.

For the remainder of the terms, the sum turns into an integral:
\begin{align}
  q_{SR}^\nu(\hm;t) &\equiv \lim_{N,j\to\infty} j^{-1}
    \sum_{l=0}^{j-\nu}\frac{1}{(1-\tau)^l}L_l^\nu(M^2)L_{j-l}^{-\nu}(-X_k^2)\nonumber\\
   &= \left(\frac \hx \hm\right)^\nu
    \int_0^1 dr\, e^{\frac 1 2r t\hd^2}\left(\frac r {1-r}\right)^{\nu/2}
    I_\nu(\sqrt{rt}\hm)I_\nu(\sqrt{(1-r)t}\hx)
    \ .
\end{align}
The final scaling limit of the new polynomials it then the sum
\be
  q_S^\nu(\hm;t)\equiv \lim_{N,j\to\infty} \frac{(-)^j}{j!j}q_j^\nu(M^2)
  = q_{SR}^\nu(\hm;t)+q_{SA}^\nu(\hm;t)\ .
\ee
Let us give the limiting polynomials for the first few topologies
$\nu=0,1,2$ as examples:
\bea
q_S^{\nu=0}(\hm;t)&=&  q_{SR}^{\nu=0}(\hm;t)= \int_0^1dr \,
e^{\frac 1 2r t\hd^2}
    I_0(\sqrt{rt}\hm)I_0(\sqrt{(1-r)t}\hx)\ ,\\
q_S^{\nu=1}(\hm;t)&=&\frac{\hx}{ \hm}
\int_0^1 dr\, e^{\frac 1 2r t\hd^2}\left(\frac r {1-r}\right)^{1/2}
    I_1(\sqrt{rt}\hm)I_1(\sqrt{(1-r)t}\hx)
+ e^{\frac 1 2 t\hd^2}\frac{2}{\sqrt{t}\hm}I_1(\sqrt{t}\hm),\\
q_S^{\nu=2}(\hm;t)&=& \left(\frac \hx \hm\right)^2
    \int_0^1 dr\, e^{\frac 1 2r t\hd^2}\frac r {1-r}
    I_2(\sqrt{rt}\hm)I_2(\sqrt{(1-r)t}\hx)\nonumber\\
&&+\ e^{\frac 1 2 t\hd^2}\left( \frac{2\hd^2+\hx^2}{\hm^2}I_2(\sqrt{t}\hm)
+\frac{2}{\sqrt{t}\hm}I_1(\sqrt{t}\hm)\right).
\eea
For comparison see eq. (\ref{Qasymp}) for the limiting polynomials
of the previous approach \cite{AD08} for the first eigenvalue extended to
$\nu\geq0$.
Because of technical reasons the anomalous terms appear there already at $\nu=0$.

The last building block we need is the new kernel $\kerq^\nu(M^{\prime\,2},M^2)$, eq. (\ref{Knudef}).
It contains both the regular and anomalous terms
$q_{S}^\nu=q_{SR}^\nu+ q_{SA}^\nu$ inside the integral
(originating from the sum), but no further complications occur.
Hence we have the scaling limit
\be
  \kerq_S^\nu(\hm',\hm) \equiv \lim_{N\to\infty}
    \frac 1 {N^2}\kerq_N^\nu(M^{\prime\,2},M^2) =
  \frac 2 {\hm'} \int_0^1 dt\, \sqrt t I_1(\sqrt t \hm')q_{S}^\nu(\hm;t)\, .
\ee
Using these building blocks we can calculate the main quantity of interest, the
scaling limit of the
eigenvalue distributions, defined as
\be
p_{S\, k}^{(N_f,\,\nu)}(\hx_k) \equiv \int_0^{\hx_k} d\hx_1 \int_{\hx_1}^{\hx_k} d\hx_2\ldots
 \int_{\hx_{k-2}}^{\hx_k} d\hx_{k-1}\
\Omega_{S\, \nu}^{(N_f)}(\hx_1, \ldots, \hx_k)  \ ,
\label{pskdef}
\ee
with
\be
\Omega_{S\, \nu}^{(N_f)}(\hx_1, \ldots, \hx_k)
\equiv \lim_{N\to\infty}(2N)^{-k}
\Omega_{\nu}^{(N_f)}\left(x_1=\frac {\hx_1}{2N}, \ldots, x_k=\frac {\hx_k}{2N}\right)\ .
\ee

\subsection{Example partial quenching}
As pointed out in \cite{AD}, there is no $\hd$ dependence in the fully quenched
($N_1=N_2=0$) theory. The simplest case to consider is then also
what is probably the physically most relevant situation, namely
the partially quenched case, where $N_1=0$. On the lattice gauge theory side,
this corresponds to generating
the gauge configurations using $N_2$ dynamical quarks with zero
chemical potential,  and then looking at the spectrum of the Dirac operator ${\cal D}_1$
with $\hmu_1 = -\hd$.

In the simplest case $N_2 = 1$, $\nu = 0$ and $k=1$ (case $ii$),
the matrix $\kermat$ is empty,
and \eqref{intfinalbigN2} is just a $2\times 2$ determinant of
polynomials, with $M_1^{\prime}=0$.
The two columns are degenerate, so we perform a Taylor-expansion in $t$.
With the abbreviation $\hm = \hm_{f_2=1}$, the limiting distribution is then found to be
\begin{align}
  p_{S\, 1}^{(0+1,0)}(\hx_1) &= -\frac{e^{-\frac 1 2 \hd^2}}{2I_0(\hm)}
    \hx_1 e^{-\frac 1 4\hx_1^2}
    \begin{vmatrix}
      q_S^0(\hm;t=1) &\partial_t q_S^0(\hm;t)\bigr|_{t=1}\\
      1            &0
    \end{vmatrix}\\
    &= \frac{\hx_1 e^{-\frac 1 2 \hd^2-\frac 1 4\hx_1^2}}{2I_0(\hm)}
     \partial_t q_S^0(\hm;t)\bigr|_{t=1}\ ,
\end{align}
in agreement (after partial integration) with the expression in \cite{AD}.
By introducing primed masses, and changing the index of the (scaling limit
of the) new polynomials $q_S$ and the partition function, we obtain the
higher topology distributions which are new.

For $\nu=1$ (case $ii$),
we can set $\epsilon_1=0$ in eq. (\ref{defmf1shift})
from the beginning, and we find
\be
  p_{S\, 1}^{(0+1,1)}(\hx_1) = -\frac{\hm\hx_1^3e^{-\frac 1 4\hx_1^2-\frac 1 2 \hd^2}}
  {16I_1(\hm)}
    \begin{vmatrix}
      \kerq_S^1(\hx_1,\hm) &q_S^1(\hm;t=1)\\
      \kerL_S(\hx_1,0)     &1
    \end{vmatrix}\ ,
\ee
where we note that the singularities of $\kerL(\hm_1',\hm_2')$ at $\hm_2'\to\hm_1'$
and $\hm_2'\to 0$ are removable.
Increasing the topology further the primed masses become degenerate, so we take
the scaling limit with the (scaled) $\epsilon_j$'s to be finite, and then let $\epsilon_j\to 0$.
The Vandermonde $\Delta_{\nu}(\{(i\epsilon)^2\})$ of \eqref{intfinal} ensures that this limit
is nontrivial. We then find for $\nu=2$ (case $i$),
\be
  p_{S\, 1}^{(0+1,2)}(\hx_1) = -\frac{\hm^2\hx_1^4e^{-\frac 1 4\hx_1^2-\frac 1 2 \hd^2}}
  {64I_2(\hm)}
    \begin{vmatrix}
      \kerq_S^2(\hx_1,\hm) &\partial_{\hm'}\kerq_S^2(\hx_1,\hm)\\
      \kerL_S(\hx_1,0)     &\partial_{\hm'}\kerL_S(\hx_1,0)
    \end{vmatrix}\ ,
\ee
and  for $\nu=3$ (case $iii$),
\be
  p_{S\, 1}^{(0+1,3)}(\hx_1) = -\frac{\hm^3\hx_1^4e^{-\frac 1 4\hx_1^2-\frac 1 2 \hd^2}}
  {64I_3(\hm)}
    \begin{vmatrix}
      \kerq_S^3(\hx_1,\hm)      &\partial_{\hm'}\kerq_S^3(\hx_1,\hm)
      &\partial_{\hm'}^2\kerq_S^3(\hx_1,\hm)\\
      \kerL_S(\hx_1,0)          &\partial_{\hm'}\kerL_S(\hx_1,0)
      &\partial_{\hm'}^2\kerL_S(\hx_1,0)\\
      \frac 1 {\hx_1}I_1(\hx_1)
&
\frac 1 {\hx_1}I_2(\hx_1)
&\partial_{\hx_1}[\frac 1  {\hx_1}I_2(\hx_1)]
    \end{vmatrix}\ ,
\ee
where by $\partial_{\hm'}$ we mean the derivative with respect to the first argument (which,
incidentally, is not the same as $\partial_{\hx_1}$).
Our new results for $\nu=1,2$ and $3$ are illustrated in figure
\ref{fig:pnu123Nf0+1}.
For alternative
expressions in an equivalent formulation see appendix  \ref{oldnugen}.
The first eigenvalue distribution alone is in general not sufficient to fit both
low-energy constants (LEC) $\Sigma$ and $F_\pi$, see the discussion in \cite{CT3}. The benefit of computing the first eigenvalue distribution at higher topology here is to be able to fit both LEC independently from different lattice configurations, where typically $\nu=1$ offers  better statistics than $\nu=0$.
\begin{figure*}[h]
  \unitlength1.0cm
  \epsfig{file=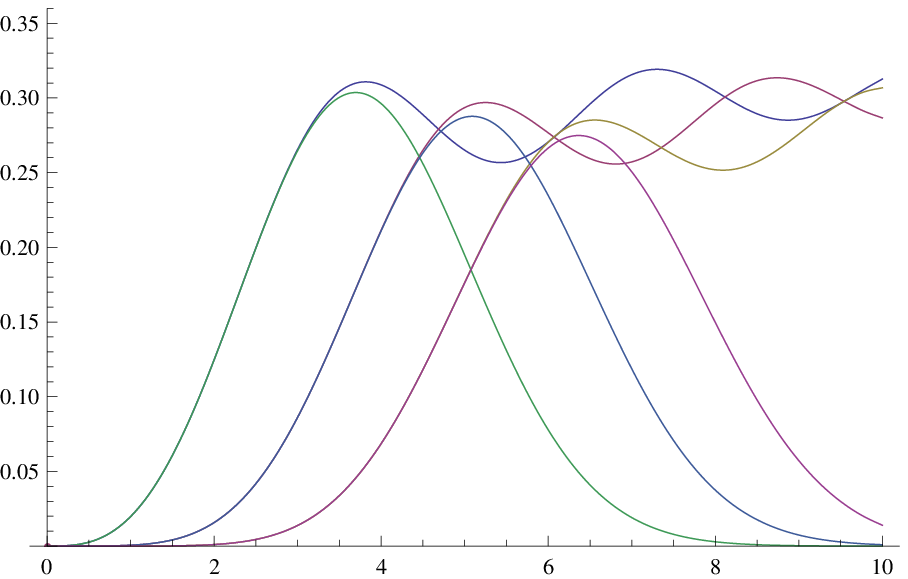,clip=,width=8cm}
    \epsfig{file=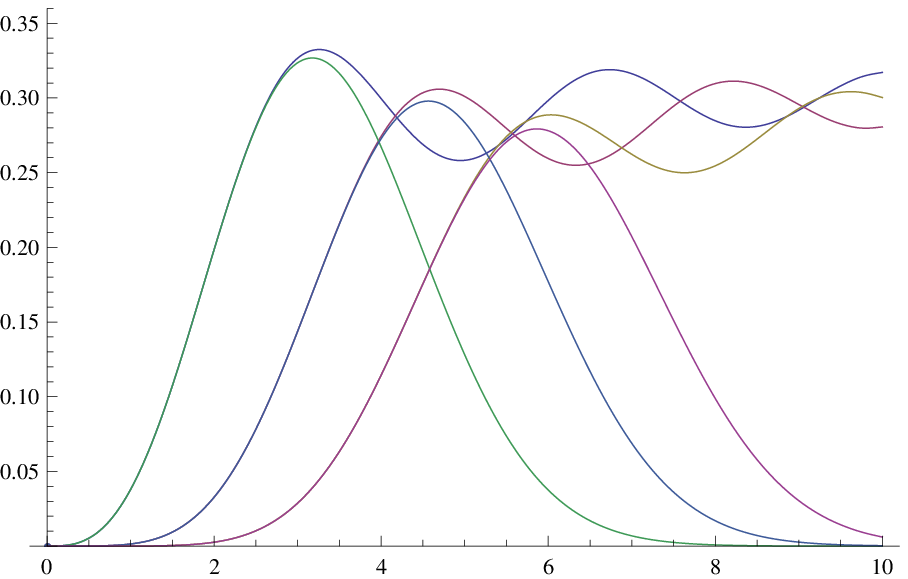,clip=,width=8cm}
  \caption{
    \label{fig:pnu123Nf0+1}
The first eigenvalue in
the partially quenched case with $N_1=0$ and $N_2=1$ at mass
$\hm=3$,
shown for 3
different topologies and 2 different values of $\hat{\delta}=1$ (left
figure) and $\hat{\delta}=3$ (right figure)
The left, middle and right curves correspond to topology
$\nu=1,2$ and $3$, respectively. The effect of the increasing number
of exact zero-eigenvalues can be seen nicely.
We also show the corresponding spectral density from \cite{ADOS}
for comparison (see eq. (\ref{rho2MM01}) in appendix \ref{oldnugen}).
The first eigenvalue always follows the density almost up to its
first maximum for all parameter values, which is an important
consistency check.
}
\end{figure*}

Continuing with the new expressions for the $k$th
lowest eigenvalues for $N_f=0+1$ with $k>1$, we will, for simplicity,
focus on $\nu=0$.
Results at higher topology are easily
obtained following the same approach as above. For $k=2$ and $k=3$
the joint probability distribution is found to be respectively
\be
  \Omega_{S\, 0}^{(0+1)}(\hx_1,\hx_2) =
  -\frac{e^{-\frac 1 2 \hd^2}}{16I_0(\hm)}\hx_1\hx_2(\hx_2^2-\hx_1^2)e^{-\frac 1 4 \hx_2^2}
  \begin{vmatrix}
    \kerq_S^0(\hx_{21},\hm) &q_S^0(\hm;t=1) &\partial_t q_S^0(\hm;t)\bigr|_{t=1}\\
    \kerL_S(\hx_{21},\hx_{21}) &\frac 2 {\hx_{21}} I_1(\hx_{21}) &I_2(\hx_{21})\\
    \kerL_S(\hx_{21},0) &1 &0
  \end{vmatrix}\, .
\ee
and
\begin{multline}
  \Omega_{S\, 0}^{(0+1)}(\hx_1,\hx_2,\hx_3) =
  -\frac{e^{-\frac 1 2 \hd^2}}{128I_0(\hm)}\hx_1\hx_2\hx_3(\hx_3^2-\hx_1^2)(\hx_3^2-\hx_2^2)
  e^{-\frac 1 4 \hx_3^2}\\
  \times
  \begin{vmatrix}
    \kerq_S^0(\hx_{31},\hm) &\kerq_S^0(\hx_{32}) &q_S^0(\hm;t=1) &\partial_t q_S^0(\hm;t)\bigr|_{t=1}\\
    \kerL_S(\hx_{31},\hx_{31}) &\kerL_S(\hx_{32},\hx_{31}) &\frac 2 {\hx_{31}}I_1(\hx_{31}) &I_2(\hx_{31})\\
    \kerL_S(\hx_{31},\hx_{32}) &\kerL_S(\hx_{32},\hx_{32}) &\frac 2 {\hx_{32}}I_1(\hx_{32}) &I_2(\hx_{32})\\
    \kerL_S(\hx_{31},0) &\kerL_S(\hx_{32},0) &1 &0
  \end{vmatrix}\, .
\end{multline}
Here we have used the shorthand
$\hx_{ij}\equiv\sqrt{\hx_i^2-\hx_j^2}$. The eigenvalue distributions
then follow by applying eq. (\ref{pskdef}).
\begin{figure*}[h]
  \unitlength1.0cm
  \epsfig{file=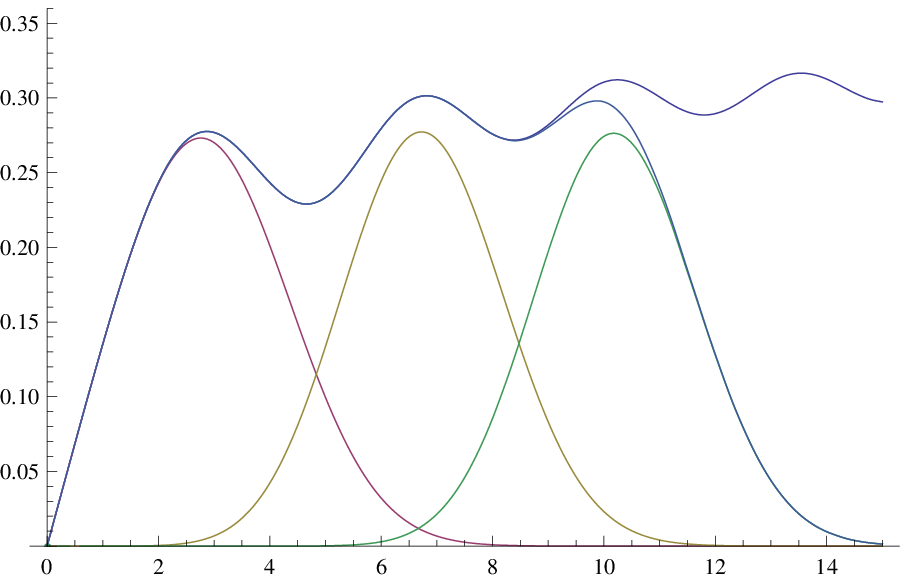,clip=,width=8cm}
    \epsfig{file=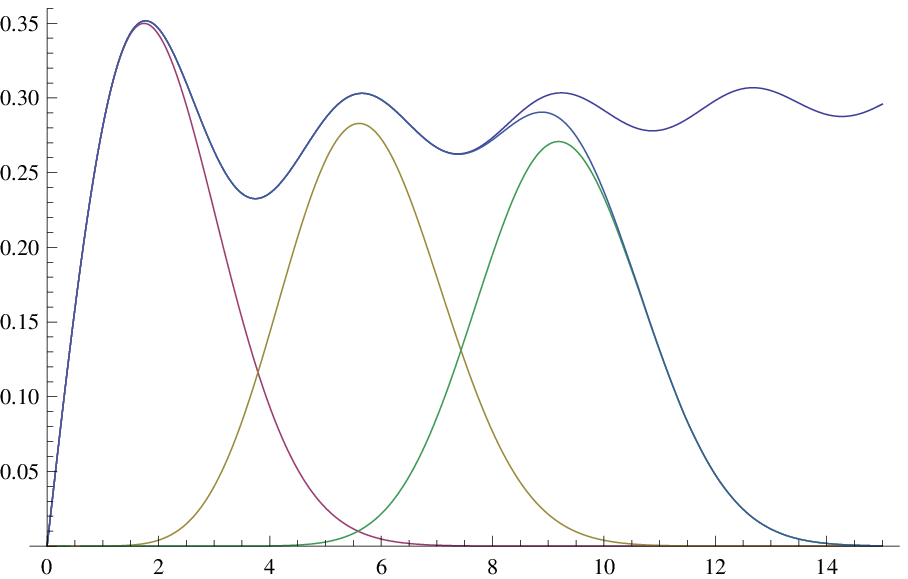,clip=,width=8cm}
  \caption{
    \label{fig:p123nu0Nf0+2}
The first, second and third eigenvalue $k=1,2,3$ in
the partially quenched case with $N_1=0$ and $N_2=2$ at masses
$\hm_1=3$ and $\hm_1=4$, all at topology $\nu=0$.
We display 2 different values of $\hat{\delta}=1$ (left
figure) and $\hat{\delta}=3$ (right figure)
to illustrate the influence of chemical potential.
The left, middle and right curves correspond to
$k=1,2$ and $3$, respectively.
Also shown is the corresponding spectral density from
\cite{ADOS} (see eq. (\ref{rho2MM02}) in appendix \ref{oldnugen}). It is
compared to the sum of the first three eigenvalue distributions, following
the density identically up to the third eigenvalue for
all parameter values.
}
\end{figure*}

The extension of these results to higher values of $N_2$ is straightforward.
We will restrict ourselves here to $N_2=2$ and $\nu=0$. Defining the normalisation
constant
\newcommand{\Zzeroptwo}{\mathcal N^{(0+2)}}
\be
  \Zzeroptwo = e^{\hd^2} \begin{vmatrix}
    I_0(\hm_1) &\hm_1 I_1(\hm_1)\\
    I_0(\hm_2) &\hm_2 I_1(\hm_2)
  \end{vmatrix}\ ,
\ee
we have
\be
  p_{S\, 1}^{(0+2,0)}(\hx_1) = \frac{\hx_1e^{-\frac 1 4\hx_1^2}}
  {\Zzeroptwo} \begin{vmatrix}
    \partial_t q_S^0(\hm_1;t)\bigr|_{t=1} &\partial_t^2 q_S^0(\hm_1;t)\bigr|_{t=1}\\
    \partial_t q_S^0(\hm_2;t)\bigr|_{t=1} &\partial_t^2 q_S^0(\hm_2;t)\bigr|_{t=1}
  \end{vmatrix}\ .
\ee
Note that an alternative representation for the integral of this
quantity, it cumulative distribution, was given in eq. (4.44) in \cite{AD08}.

For $k=2$ and $k=3$ we obtain the following new expressions
\begin{multline}
  \Omega_{S\, 0}^{(0+2)}(\hx_1,\hx_2) = -\frac{\hx_1\hx_2(\hx_2^2-\hx_1^2)e^{-\frac 1 4\hx_2^2}}
  {8\Zzeroptwo}\\ \times \begin{vmatrix}
    \kerq_S^0(\hx_{21},\hm_1)  &q_S^0(\hm_1;t=1)                &\partial_t q_S^0(\hm_1;t)\bigr|_{t=1} &\partial_t^2 q_S^0(\hm_1;t)\bigr|_{t=1}\\
    \kerq_S^0(\hx_{21},\hm_2)  &q_S^0(\hm_2;t=1)                &\partial_t q_S^0(\hm_2;t)\bigr|_{t=1} &\partial_t^2 q_S^0(\hm_2;t)\bigr|_{t=1}\\
    \kerL_S(\hx_{21},\hx_{21}) &\frac 2 {\hx_{21}}I_1(\hx_{21}) &I_2(\hx_{21})                         &\frac{\hx_{21}}{2}I_3(\hx_{21})\\
    \kerL_S(\hx_{21},0)        &1                               &0                                     &0
  \end{vmatrix}\ ,
\end{multline}
and
\begin{multline}
  \Omega_{S\, 0}^{(0+2)}(\hx_1,\hx_2,\hx_3) = \frac{\hx_1\hx_2\hx_2(\hx_3^2-\hx_1^2)(\hx_3^2-\hx_2^2)e^{-\frac 1 4\hx_3^2}}
  {2^6\Zzeroptwo}\\ \times \begin{vmatrix}
    \kerq_S^0(\hx_{31},\hm_1)  &\kerq_S^0(\hx_{32},\hm_1)  &q_S^0(\hm_1;t=1)                &\partial_t q_S^0(\hm_1;t)\bigr|_{t=1} &\partial_t^2 q_S^0(\hm_1;t)\bigr|_{t=1}\\
    \kerq_S^0(\hx_{31},\hm_2)  &\kerq_S^0(\hx_{32},\hm_2)  &q_S^0(\hm_2;t=1)                &\partial_t q_S^0(\hm_2;t)\bigr|_{t=1} &\partial_t^2 q_S^0(\hm_2;t)\bigr|_{t=1}\\
    \kerL_S(\hx_{31},\hx_{31}) &\kerL_S(\hx_{32},\hx_{31}) &\frac 2 {\hx_{31}}I_1(\hx_{31}) &I_2(\hx_{31})                         &\frac{\hx_{31}}{2}I_3(\hx_{31})\\
    \kerL_S(\hx_{31},\hx_{32}) &\kerL_S(\hx_{32},\hx_{32}) &\frac 2 {\hx_{32}}I_1(\hx_{32}) &I_2(\hx_{32})                         &\frac{\hx_{32}}{2}I_3(\hx_{32})\\
    \kerL_S(\hx_{31},0)        &\kerL_S(\hx_{32},0)        &1                               &0                                     &0
  \end{vmatrix}\ ,
\end{multline}
after inserting them into  eq. (\ref{pskdef}).
We illustrate the distributions of individual eigenvalues following from these equations in
figure \ref{fig:p123nu0Nf0+2} for non-degenerate masses.
In the case of equal masses, two rows become degenerate in both the numerator
and denominator, necessitating a Taylor expansion.
Figure \ref{fig:p123nu0Nf0+2} offers a further graphical consistency check for our results, when comparing the sum of the individual eigenvalues to the actual density. The two curves nicely agree almost up to the third local maximum for all parameter values.
%

\subsection{Two light flavours}

For $N_1=N_2=1$, $\nu = 0$ the distribution of the first eigenvalue is given
by
\newcommand{\Zonepone}{\mathcal N^{(1+1)}}
\be
  p_{S\, 1}^{(1+1,\,0)}(\hx_1) = -\frac {\hx_1(\hm_1^2+\hx_1^2)e^{-\frac 1 4\hx_1^2}}
   {8\Zonepone}
  \begin{vmatrix}
    \kerq_S^0(\hm',\hm_2) &q_S^0(\hm_2;t=1)\\
    \kerL_S(\hm',0)       &1
  \end{vmatrix}\, ,
\ee
\be
  \Zonepone = \int_0^1dt\, e^{\frac 1 2 t\hat{\delta}^2}
    I_0(\sqrt t \hm_1)I_0(\sqrt t \hm_2) \ ,
\ee
with $\hm_1 = \hm_{f_1=1}$, $\hm_2 = \hm_{f_1=2}$ and $\hm' =
\sqrt{\hm_1^2+\hx_k^2}$. For its cumulative distribution see
eq. (4.33) in \cite{AD08}.

\begin{figure*}[h]
  \unitlength1.0cm
  \epsfig{file=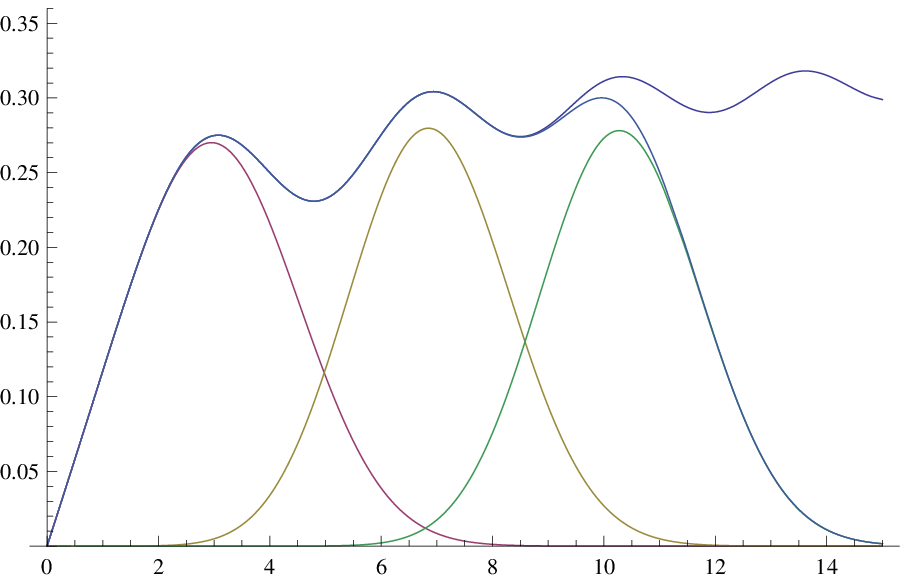,clip=,width=8cm}
    \epsfig{file=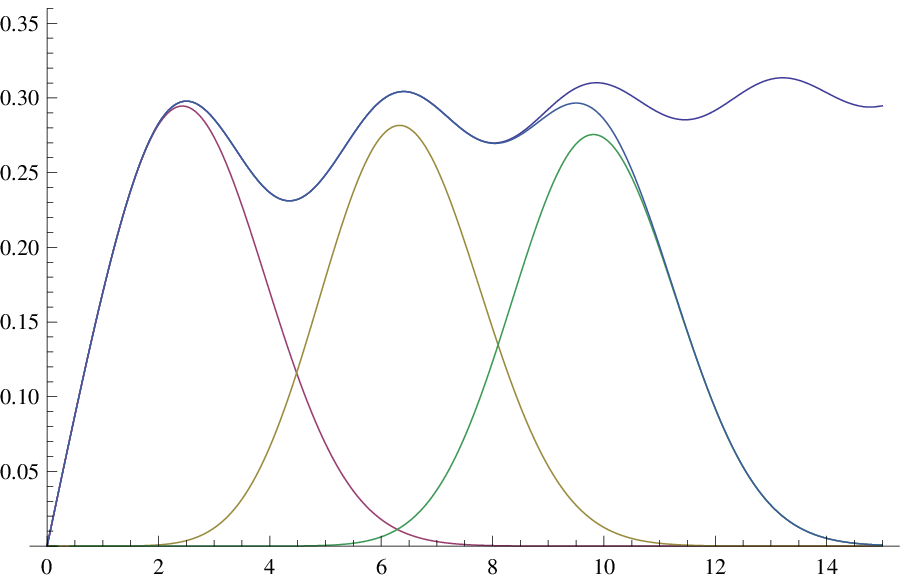,clip=,width=8cm}
  \caption{
    \label{fig:p123nu0Nf1+1}
The first, second and third eigenvalue $k=1,2,3$
for two light flavours with $N_1=1$ at mass $\hm_1=3$
and $N_2=1$ at mass $\hm_1=4$, all at topology $\nu=0$.
We display 2 different values of $\hat{\delta}=1$ (left
figure) and $\hat{\delta}=3$ (right figure) 
to illustrate the influence of chemical potential.
The left, middle and right curves correspond to
$k=1,2$ and $3$, respectively.
The corresponding spectral density is taken from from
\cite{ADOS}(see eq. (\ref{rho2MMN11}) in appendix \ref{oldnugen}). Again we find an excellent agreement between the density
and the sum of the three individual eigenvalues for all parameter
values chosen.
}
\end{figure*}

The expressions for distribution of the second and third eigenvalue which are new results follow in a similar fashion from
\be
  \Omega_{S\, 0}^{(1+1)}(\hx_1,\hx_2) =
  \frac {\hx_1\hx_2(\hx_2^2-\hx_1^2)(\hm_1^2+\hx_2^2)e^{-\frac 1 4\hx_2^2}}
   {64\Zonepone}
  \begin{vmatrix}
    \kerq_S^0(\hm',\hm_2)  &\kerq_S^0(\hx_{21},\hm_2)  &q_S^0(\hm_2t=1)\\
    \kerL_S(\hm',\hx_{21}) &\kerL_S(\hx_{21},\hx_{21}) &\frac 2 {\hx_{21}}I_1(\hx_{21})\\
    \kerL_S(\hm',0)        &\kerL_S(\hx_{21},0)            &1
  \end{vmatrix}\, ,
\ee
and
\begin{multline}
  \Omega_{S\, 0}^{(1+1)}(\hx_1,\hx_2,\hx_3) =
  -\frac {\hx_1\hx_2\hx_3(\hx_3^2-\hx_1^2)(\hx_3^2-\hx_2^2)(\hm_1^2+\hx_3^2)
    e^{-\frac 1 4\hx_3^2}}
  {2^9\Zonepone}\\
  \times
  \begin{vmatrix}
    \kerq_S^0(\hm',\hm_2)  &\kerq_S^0(\hx_{31},\hm_2)  &\kerq_S^0(\hx_{32},\hm_2)  &q_S^0(\hm_2;t=1)\\
    \kerL_S(\hm',\hx_{31}) &\kerL_S(\hx_{31},\hx_{31}) &\kerL_S(\hx_{32},\hx_{31}) &\frac 2 {\hx_{31}}I_1(\hx_{31})\\
    \kerL_S(\hm',\hx_{32}) &\kerL_S(\hx_{31},\hx_{32}) &\kerL_S(\hx_{32},\hx_{32}) &\frac 2 {\hx_{32}}I_1(\hx_{32})\\
    \kerL_S(\hm',0)        &\kerL_S(\hx_{31},0)        &\kerL_S(\hx_{32},0)        &1
  \end{vmatrix}\, .
\end{multline}
The corresponding figures are shown in figure \ref{fig:p123nu0Nf1+1}.

Both figures \ref{fig:p123nu0Nf0+2} and \ref{fig:p123nu0Nf1+1} illustrate the influence of chemical potential and thus the possibility to determine $F_\pi$ from individual eigenvalue distributions. In particular the shape of the first eigenvalue, but also the positions of the maxima of the second and third eigenvalue change considerably when comparing the left and right plots, in particular in the partially quenched setting in figure \ref{fig:p123nu0Nf0+2}. Let us briefly compare to how $\Sigma$ and $F_\pi$ were determined from lattice data in ref. \cite{CT3}. In order to avoid two-parameter fits, there $\Sigma$ was first determined at $\mu_{1,2}=0$ from the first eigenvalue alone. Then, the two-point density correlation function $\rho(\hx,\hy)$ of finding one $\hx$- and one $\hy$-eigenvalue was expanded for small $\hat{\delta}$, and $F_\pi$ was fitted to the resulting Gaussian repulsion between $\hx$ and $\hy$.
We would expect that the detailed knowledge of the shape and spacing of several individual eigenvalues combined with independent measurements at different topologies could provide an alternative way to determine both low-energy constants.
%

\sect{Conclusions}\label{conc}

We have derived new analytical expressions for the probability
distribution of the $k$th lowest Dirac operator eigenvalue in QCD
with three (or more) colours
with spontaneous breaking of chiral symmetry. Extending
earlier results we have shown how these distributions
become modified when the Dirac operator couples to imaginary
chemical potential, for arbitrary topology and including the partially quenched case.
Because the deformation parameter of the spectrum is
$\hat{\mu} = \mu F_{\pi}\sqrt{V}$ (where $\mu$ is the externally
supplied isospin chemical potential), one can use this to
extract the magnitude of $F_{\pi}$ through lattice gauge theory
simulations. This been already successfully done for the first
eigenvalue at $\nu=0$, after including first order finite-volume
corrections via $F_{\pi\,ef\!f}$.
Our results thus offer further independent checks for higher topology
or higher eigenvalues, given they remain in the $\epsilon$-regime.

There is a clear sensitivity to $F_{\pi}$ in these individual eigenvalue
distributions, but it requires quark masses to be quite small.
Because present-day lattice gauge configurations may tend to be available
for relatively large masses, far from the scale of eigenvalues considered
here, the eigenvalue distributions will look close to quenched
and there is then little variation in the distributions
as $\mu$ is introduced.
For this reason it would be extremely helpful if one could derive analogous
analytical expressions for mixed or conditional individual eigenvalue distributions, say to find simultaneously the first eigenvalue of ${\cal D}_1$ at $x$ and the first eigenvalue of ${\cal D}_2$ at $y$.
In the same way as the mixed spectral two-point density correlation function $\rho(x,y)$
considered previously in the literature, that develops a
delta-function in the $\hat{\mu}\to0$ limit,
this quantity
should lead to a more dramatic
numerical signal as chemical potential is turned on and thus to
an easier way of measuring $F_{\pi}$ by means of such a technique. It is an
open challenge to find new mathematical tricks that will be needed
to perform this extension of our results.

\vspace{0.5cm}
{\sc Acknowledgements}:~ We are indebted to Poul Damgaard for
many discussions and an early collaboration on this topic. We would
also like to thank Christoph Lehner and Tilo Wettig for discussions
and correspondence regarding their work.

This work was supported partly by
EPSRC grant EP/D031613/1, EU network ENRAGE MRTN-CT-2004-005616 and the
Niels Bohr Foundation (G.A.). The Niels Bohr Institute is also thanked
for its warm hospitality throughout several visits (G.A.).

\begin{appendix}

\sect{Identity for Laguerre Polynomials}
\label{Qid}
In this Appendix we derive the form of the new polynomials given in
eq. (\ref{qfinal}). They first appear in the manipulations leading to
eq. (\ref{Lfinal}) in the following form
 \bea
q_n^\nu(M_2^2)&\equiv&
\sum_{j=0}^n \frac{(-)^{n+j}n!(n+1)!}{(n-j)!j!(j+1)!}
\sum_{l=0}^j\frac{\tau^l}{(1-\tau)^{l}}
\hat{L}_l^\nu\left(\frac{M^2_2}{\tau}\right)(-X_k^2)^{j-l}
{j \choose l}\nn\\
&=& (-)^n n!
\sum_{l=0}^n \frac{\tau^l}{(1-\tau)^{l}}
{L}_l^\nu\left(\frac{M^2_2}{\tau}\right)L_{n-l}^{l+1}(-X_k^2)\nn\\
&=& (-)^n n!\sum_{l=0}^n L_{n-l}^{l+1}(-X_k^2)\sum_{j=0}^l
\frac{(-)^{j+l}(l+\nu)!}{(l-j)!(j+\nu)!(1-\tau)^j}L_j^\nu(M_2^2)\nn\\
&=& (-)^n n!
\sum_{j=0}^n
\frac{1}{(1-\tau)^j}L_j^\nu(M_2^2) \sum_{i=0}^{n-j}
\frac{(-)^i(i+j+\nu)!}{(j+\nu)!i!}L_{n-j-i}^{i+j+1}(-X_k^2)\ .
\label{qid}
\eea
The first line originates from generating polynomials $\hat{L}_n^1$ in the
lower rows in eq. (\ref{Lfinal}).
Next we have swapped the sums, $\sum_{j=0}^n\sum_{l=0}^j\to
\sum_{l=0}^n\sum_{j=l}^n$, so that the sum in $-X_k^2$ gives
$L_{n-l}^{l+1}(-X_k^2)$. Then we use
the following identity:
\be
L_{n}^\nu(zw)=\sum_{j=0}^{n}
\frac{(n+\nu)!}{(n-j)!(\nu+j)!} w^j(1-w)^{n-j}
L_j^\nu\left(z\right) \ .
\label{Lid2}
\ee
to eliminate the argument $1/\tau$ from the first Laguerre polynomial,
choosing $z=M_2^2$ and $w=1/\tau$. The newly appearing sum is
swapped again to $\sum_{j=0}^n\sum_{l=j}^n$. The remaining sum can be
simplified with the help of the identity
\be
L_{m}^{-\nu}(x)\ =\
\sum_{i=0}^{m}\frac{(-)^i(n+\nu-m+i)!}{(n+\nu-m)!i!}L_{m-i}^{n-m+i+1}(x) ~,
\label{LidII}
\ee
with $m=n-j$ and $x=-X_k^2$.

The polynomials $q_n^\nu(x)$ and their derivation differ slightly
from the polynomials $Q_n(x)$ in \cite{AD08}. First, we allow for non-zero
topology here, in contrast to there. Second, in the related formula to
(\ref{Lfinal}) the Laguerre polynomials $L_n^0$ have to be generated
there, instead of $L_n^1$ here.

For completeness we also define the generalization of the polynomials $Q_n(x)$ 
that are needed to complete the computations
of \cite{AD08} for non-zero topology:
\bea
Q_n^\nu(M_2^2)&\equiv&
\sum_{j=0}^n \frac{(-)^{n+j}(n!)^2}{(n-j)!(j!)^2}
\sum_{l=0}^j\frac{\tau^l}{(1-\tau)^{l}}
\hat{L}_l^\nu\left(\frac{M^2_2}{\tau}\right)(-S^2)^{j-l}
{j \choose l}\nn\\
&=& (-)^n n!
\sum_{j=0}^n
\frac{1}{(1-\tau)^j}L_j^\nu(M_2^2) L_{n-j}^{-\nu-1}(-S^2)
\ ,
\label{Qdef}
\eea
with $S^2=+N\tau c_1s^2$.
The derivation goes along exactly
the same lines as above and we skip the details here.
The single difference compared to eq. (\ref{qfinal})
is the shift in the negative index
from $-\nu\to-\nu-1$ in the final result.
In Appendix \ref{oldnugen} below we give the corresponding expressions for
the first eigenvalue from \cite{AD08} for non-zero $\nu$.


\sect{Integration Theorem over Determinants}
\label{AVgen}
In this Appendix we compute averages of the product of
two determinants of different size that both contain
different variables {\em and}
different polynomials. The Theorem we give is a slight
generalisation of Theorem 1 in \cite{AV}. It was already used in \cite{AD08}
but since the form we need here is more general we provide
a short derivation.
In particular, it generalises Dyson's Theorem, see Theorem 5.1.4 in
\cite{Mehta}, valid only for determinants of the
same size and of the same polynomials.

Suppose we have two sets of (bi)orthonormal functions (this includes the
case of polynomials orthogonal w.r.t. a given weight, since the weight
can be absorbed
into the functions) $\varphi_{j=0,1,\ldots}(z)$
and $\tP_{j=0,1,\ldots}(z)$
\be
\int dz\ \varphi_{j}(z) \tP_l(z)\ =\ \delta_{jl}\ \ \mbox{for}\ j,l=0,1,\ldots\
.
\label{OPdef}
\ee
Furthermore we assume that we have two matrices (independent of $z_1,\ldots,z_N$),
$A_{ij}$ and $B_{ij}$.
We can then compute the following average over  $N$ variables
$z_j$ with $k\geq1$:
\bea
{\cal I}\equiv
\int_{0}^{\infty}\prod_{j=1}^N dz_j\
\left|\begin{array}{ccc}
    A_{1,0}  &\cdots &A_{1,N+k-1}   \\
    \vdots   &       &\vdots        \\
    A_{k,0}  &\cdots &A_{k,N+k-1}   \\
    \varphi_0(z_1) &\cdots &\varphi_{N+k-1}(z_1)\\
    \vdots   &       &\vdots        \\
    \varphi_0(z_N) &\cdots &\varphi_{N+k-1}(z_N)
  \end{array}\right|
\left|\begin{array}{ccc}
    B_{1,0}         &\cdots &B_{1,N+k'-1}          \\
    \vdots          &       &\vdots                \\
    B_{k',0}        &\cdots &B_{k',N+k'-1}         \\
    \tilde \varphi_0(z_1) &\cdots &\tilde \varphi_{N+k'-1}(z_1)\\
    \vdots          &       &\vdots                \\
    \tilde \varphi_0(z_N) &\cdots &\tilde \varphi_{N+k'-1}(z_N)
\end{array}\right|,
\label{Th}
\eea
where without loss of generality we assume $k\geq k'$.
We need to define 4 different objects
that we refer to as kernels
\bea
&&K^I_{N+k'}(i,j) = \sum_{l=0}^{N+k'-1}A_{i,l}B_{j,l}\ ,\
    K^{II}_{N+k'}(i,z) = \sum_{l=0}^{N+k'-1}A_{i,l}\tilde \varphi_l(z)\ ,\cr
&&K^{III}_{N+k'}(z,j) = \sum_{l=0}^{N+k'-1}\varphi_l(z)B_{j,l}\ ,\
    K^{IV}_{N+k'}(z,z') = \sum_{l=0}^{N+k'-1}\varphi_l(z)\tilde \varphi_l(z')\ .
\label{Kdefs}
\eea
Only the last kernel has the self-reproducing properties required for Dyson's
Theorem to apply:
\bea
&&\int dz' K^{IV}_{N+k'}(z,z')K^{IV}_{N+k'}(z',z'')= K^{IV}_{N+k'}(z,z'')\ \ ,\ \cr
&&\int dz' w(z')K_{N+k'}^{IV}(z',z')=N+k'\ \ .
\label{Kself}
\eea
Obviously the contraction of mixed kernels leads to mixed kernels,
\bea
&&\int dz' K_{N+k'}^{II}(i,z')K_{N+k'}^{III}(z',j)= K_{N+k'}^{I}(i,j)\ \ ,\cr
&&\int dz' K_{N+k'}^{II}(i,z')K_N^{IV}(z',z)= K_{N+k'}^{II}(i,z)\ \ ,\cr
&&\int dz' K_{N+k'}^{IV}(z,z')K_{N+k'}^{III}(z',i)=K_{N+k'}^{III}(z,i)\ \ .
\eea
The result is now given in terms of the matrix
\newcommand{\Bmat}{\mathcal B}
\[
  \Bmat_{ij} = \begin{cases}
    K^{I}_{N+k'}(i,j) &i=1,\ldots,k, j=1,\ldots,k'       \\
    A_{i,j+N-1}  &i=1,\ldots,k, j=k'+1,k'+2,\ldots,k
  \end{cases}.
\]
The claim is that the integral \eqref{Th} is given by
\be
{\cal I} \ =\  (-)^{N(k-k')}N!\det[ \Bmat]\ .
  \label{ThResult}
\ee
By choosing $A_{ij}$ and $B_{ij}$ to be appropriate polynomials of the
unintegrated variables, letting $\varphi_n(z)$ and $\tilde \varphi_n(z)$ be Laguerre
polynomials times the weight factor, and
choosing $N-k$ variables $z_1,\ldots,z_N \to Z_{k+1},\ldots,Z_N$
 we arrive at the integral
of \eqref{Omdelta3}. Taking the integration norm into account, we see
that \eqref{intfinal}, \eqref{intfinalbigN2} and \eqref{intfinalsmallN2}
follow from \eqref{ThResult}.

The proof of the formula is by induction on $k$. We first consider the base
case $k = k'$, where \eqref{ThResult} is just $N!\det[ K^I(i,j)]$.
To derive this, we first transpose one of the matrices in \eqref{Th},
and then multiply them inside a common determinant. The resulting
matrix will then consist of blocks with each of the four types of
kernels. Using the self-reproducing property of $K^{VI}$ and
Dyson's Theorem \cite{Mehta}, we can carry out the integration,
yielding the stated result.

For the step $k \to k+1$, we  expand the first
determinant of \eqref{Th} in the last column:
\[
\sum_{i=1}^{k+N+1}(-)^{k+N+1+i}
    \int_{0}^{\infty}\left(\prod_{j=1}^N dz_j\right)
    (M_A)_{i,k+N+1}\det[ m_A(i)]\det[ M_B],
\]
with $M_A$ ($M_B$) the first (second) matrix of \eqref{Th}, and
$m_A(i)$ being $M_A$ with the row $i$ and column $k+N+1$ removed.
The integrands with $i > k+1$ are proportional to $\varphi_{k+N}(z_{i-k-1})$,
but since there is no corresponding $\tilde \varphi_{k+N}(z_{i-k-1})$
factor in the expansion of the $M_B$ determinant, these terms are
killed by orthogonality. Using the induction hypothesis on the remaining
terms (proportional to $A_{i,k+N}$) and comparing to the expansion
of $\det \Bmat$ in the last column, we verify \eqref{ThResult}.

\sect{Non-Zero Topology for $p_{k=1}^{(N_f,\nu)}(x)$: an  Alternative Formulation}
\label{oldnugen}

In this appendix we derive an alternative formulation of the distribution of the first eigenvalue
for $\nu>0$. It is based on an earlier paper \cite{AD08} to where we refer for more details,
explicit results were given there only for $\nu=0$.
Because the calculations in the main part of this paper are quite involved
it is useful to have an independent,
equivalent result as a cross check.

In \cite{AD08} one computes first the gap probability\footnote{It is denoted there by $E_{0,0}(s,t=0)$.},
that the interval $[0,s]$ is empty of all eigenvalues $x_i$:
\be
E_\nu^{(N_f)}(s) \equiv\frac{1}{{\cal Z}_\nu^{(N_f)}}
\int_s^\infty dx_1\ldots dx_N
\int_0^\infty dy_1 \ldots dy_N \
{\cal P}_{\nu}^{(N_f)}(\{x\},\{y\};\{m_1\},\{m_2\})\ .
\label{1gapdef}
\ee
From this the distribution of the first eigenvalue follows by differentiation,
$p_1^{(N_f,\nu)}(s)=-\partial_s E_\nu^{(N_f)}(s)$.
In steps very similar to the main body of this paper we obtain
\bea
E_\nu^{(N_f)}(s)&\sim& \frac{1}{
{\cal Z}_\nu^{(N_1+N_2)}\Delta_{N_1+\nu}(m_{f_1}^{\prime\,2})\Delta_{N_2}(m_{f_2}^2)}
e^{-N^2\tau c_1s^2}
\int_0^\infty dz_1\ldots dz_N\ e^{-N\sum_{i=1}^N \tau c_1 z_i}
\label{EfullNf}\\
&&\times
\left|
\begin{array}{lll}
\hat{L}_0(M_{1}^{\prime\,2})
& \cdots &\hat{L}_{N+N_1+\nu-1}(M_{1}^{\prime\,2})
\\
\cdots & \cdots  &\cdots\\
\hat{L}_0(M_{N_1+\nu}^{\prime\,2})
& \cdots &\hat{L}_{N+N_1+\nu-1}(M_{N_1+\nu}^{\prime\,2})
\\
\hat{L}_0(Z_1) & \cdots &
\hat{L}_{N+N_1+\nu-1}(Z_1)
\\
\cdots & \cdots  &\cdots\\
\hat{L}_0(Z_N) & \cdots &
\hat{L}_{N+N_1+\nu-1}(Z_N)
\\
\end{array}
\right|\left|
\begin{array}{lllll}
Q_0^{\nu}(M_{1}^2)
& \cdots &Q_{N+N_2-1}^{\nu}(M_{1}^2)
\\
\cdots & \cdots&\cdots\\
Q_0^{\nu}(M_{N_2}^2)
& \cdots &Q_{N+N_2-1}^{\nu}(M_{N_2}^2)
\\
\hat{L}_0(Z_1) & \cdots &
\hat{L}_{N+N_2-1}(Z_1)
\\
\cdots & \cdots &\cdots\\
\hat{L}_0(Z_N) & \cdots &
\hat{L}_{N+N_2-1}(Z_N)
\\
\end{array}
\right|,
\nn
\eea
where we have suppressed all mass independent normalisation factors.
The Laguerre polynomials $\hat{L}_k(x)=(-)^k k!L_k(x)$ in monic normalisation
contain the following masses:
\be
M_{j}^{2}\equiv -N\tau c_2m_j^2\ ,\ \ j=1,\ldots,N_2
\ee
for the flavours of type $N_2$, and for flavour $N_1$ we have $\nu$ additional masses
\bea
M_{j}^{\prime\,2}&\equiv&-N\tau c_1 m_j^{\prime\,2}
\equiv-N\tau c_1 (m_j^2+s^2)\ ,\ \ j=1,\ldots,N_1\ ,\nn\\
M_{j}^{\prime\,2} &\equiv&-N\tau c_1 m_j^{\prime\,2}
\equiv -N\tau c_1 (s^2+\epsilon_j)\ ,\ \ j=N_1+1,\ldots,N_1+\nu\ .
\eea
The degeneracy of the latter can be reinstated by setting $\epsilon_j=0$ at the
end of the calculation. Alternatively we could write out the resulting derivatives
acting on the first determinant in eq. (\ref{EfullNf}) for $\nu>1$.
Last but not least we have introduced a set of new polynomials containing now generalised
Laguerre polynomials, as defined in \eqref{Qdef}.
This completes in principle the computations
of \cite{AD08} for non-zero topology.
The derivation from the first to the last line in eq. (\ref{Qdef})
goes along the same lines as the previous appendix \ref{Qid} for the polynomials
$q_n^\nu(x)$, apart from the slightly different identity to be used:
\be
L_{n-j}^{-\nu-1}(x)\ =\
\sum_{i=0}^{n-j}\frac{(-)^i(j+\nu+i)!}{(j+\nu)!i!}L_{n-j-i}^{j+i}(x) ~.
\label{LidIII}
\ee

Applying the Theorem derived in the next appendix \ref{Sonine}
the $N$ integrals over variables $z_k$
can now be easily performed. Instead of giving the most general result, distinguishing
between different cases depending on the numbers $N_1$, $N_2$ and $\nu$, we give a few
simple examples for illustration. These can be stated in terms of the above polynomials
as well as the following kernel:
\be
{\cal K}^\nu_{N} (M_1^{\prime\,2},M_2^2) \equiv
\sum_{j=0}^{N-1} \frac{(-)^j}{j!} L_j^0(M_1^{\prime\,2})
Q_j^\nu(M_2^2)\ .
\label{KernelLQ}
\ee
Note that only the second polynomial $Q_j^\nu$ gets modified when $\nu\neq0$,
compared to \cite{AD08}. The kernel is obviously not symmetric in its arguments.

In the first example we choose $N_1=N_2=1$ to get
\bea
E_{\nu=0}^{(1+1)}(s) &\sim& \frac{e^{-N^2\tau c_1s^2}}{{\cal Z}_0^{(1+1)}(m_1;m_2)}
{\cal K}_{N+1}^{\nu=0} (M_1^{\prime\,2},M_2^2)\ ,\nn\\
E_{\nu=1}^{(1+1)}(s) &\sim&\frac{e^{-N^2\tau c_1s^2}}{{\cal Z}_1^{(1+1)}(m_1;m_2)
(m_1^{\prime\,2}-s^2)}
\left|
\begin{array}{ll}
{\cal K}_{N+1}^{\nu=1} (M_1^{\prime\,2},M_2^2) &{\cal K}_{N+1}^{\nu=1} (S^{2},M_2^2)\\
L_{N+1}^0(M_1^{\prime\,2}) & L_{N+1}^0(S^2)\\
\end{array}
\right|
\  ,
\eea
etc., with more rows with Laguerre polynomials for higher $\nu$.

The second example which is probably most relevant for applications is partially quenched, with $N_1=0$ and $N_2=2$.
Here the size of the determinant does not grow immediately as the masses $m_{1,2}$ of flavour $N_2$ get paired with
those generated by topology, $\nu=1,2$ in our examples:
\bea
E_{\nu=0}^{(0+2)}(s) &\sim&\frac{e^{-N^2\tau c_1s^2}}{{\cal Z}_0^{(0+2)}(m_{1,2})(m_1^{2}-m_2^2)}
\left|
\begin{array}{ll}
{Q}_{N}^{\nu=0} (M_1^{2}) &{Q}_{N}^{\nu=0} (M_2^2)\\
Q_{N+1}^{\nu=0}(M_1^{2}) & Q_{N+1}^{\nu=0}(M_2^2)\\
\end{array}
\right|
\ ,\nn\\
E_{\nu=1}^{(0+2)}(s) &\sim&\frac{e^{-N^2\tau c_1s^2}}{{\cal Z}_1^{(0+2)}(m_{1,2})(m_1^{2}-m_2^2)}
\left|
\begin{array}{ll}
{\cal K}_{N+1}^{\nu=1} (S^2,M_1^2) &{\cal K}_{N+1}^{\nu=1} (S^{2},M_2^2)\\
Q_{N+1}^{\nu=1}(M_1^{2}) & Q_{N+1}^{\nu=1}(M_2^2)\\
\end{array}
\right|
\ ,\nn\\
E_{\nu=2}^{(0+2)}(s) &\sim&\frac{e^{-N^2\tau c_1s^2}}{{\cal Z}_2^{(0+2)}(m_{1,2})
(m_1^{2}-m_2^2)2s}
\left|
\begin{array}{ll}
{\cal K}_{N+1}^{\nu=2} (S^2,M_1^2) &{\cal K}_{N+1}^{\nu=2} (S^{2},M_2^2)\\
\partial_s{\cal K}_{N+1}^{\nu=2} (S^2,M_1^2) &\partial_s{\cal K}_{N+1}^{\nu=2} (S^{2},M_2^2)\\
\end{array}
\right|
.
\eea
In all cases one further differentiation with respect to $s$ yields the
distribution of the first eigenvalue.

\subsection{The large-$N$ limit}

We will sketch here first how to take the microscopic large-$N$ limit of
the main building blocks: the generalised Laguerre polynomials $L_j^{\pm\alpha}$,
the new polynomials $Q_j^\nu$ and the kernel ${\cal K}_N^\nu$. Then we will give
two explicit examples for the distribution of the first eigenvalue at $\nu>0$.

The scaling limit is exactly as described in the main body of this paper
(and in \cite{AD08}), so we can be brief.
We begin with the new polynomials $Q_j^\nu$. The sum can be replaced by an
integral except for
a few terms, and we need the following ingredients in the limit $N\to\infty$
together with $\mu_{1,2}\to0$
\bea
&&\lim_{N,j,k\to\infty} (1-\tau)^{-k}= \exp\left[\frac12 rt\hd^2 \right]
\ \ \mbox{where}\ \
t\equiv j/N, \ \ r\equiv k/j\ , \nn\\
&& \lim_{N\to\infty} \tau c_{1,2}=1\ .
\label{scale2}
\eea
For the Laguerre polynomials the following scaling holds:
\bea
&&\lim_{N,j\to\infty}L_{j}^\nu(M^{2})
\ =\ N^\nu(rt)^{\nu/2}(2/\hm)^\nu I_{\nu}(\hm\sqrt{rt})\ ,
\label{Lasymp}\\
&&\lim_{N,j,k\to\infty} L_{j-k>\nu}^{-\nu-1}(-S^2)\ =\
N^{-\nu-1}(t(1-r))^{-(\nu+1)/2}(\hs/2)^{\nu+1}I_{\nu+1}(\hs\sqrt{1-r})\ .
\eea
In comparison to $\nu=0$, where we had to split off the $s$-independent part $L_0^{-1}=1$
from the sum in eq. (\ref{Qdef}), we now have to separate a total number of $\nu+1$ terms
 $L_{0,1,\ldots,\nu}^{-\nu-1}$ from the sum. In the large-$N$ limit these terms will seem to be
of higher order, but after some cancelations taking place they will give a contribution of
the same order as the sum itself.

To illustrate this we give the first two examples
(including the known $\nu=0$ case) for the limiting new polynomials
$\lim_{j\to\infty}\frac{(-)^j}{j!} Q_j^\nu(M^2)\equiv Q_S^\nu(\hm;t)$:
\bea
Q_S^{\nu=0}(\hm;t)
&=& \frac{\hs}{2}\int_0^1dr\ \frac{\sqrt{t}\ e^{\frac12 rt \hd^2}}{\sqrt{1-r}} I_0(\hm\sqrt{rt})
I_{1}(\hs\sqrt{(1-r)t})+\ e^{\frac12 t \hd^2}
I_0(\sqrt{t}\ \hm)
\ ,\label{Qasymp}\\
Q_S^{\nu=1}(\hm;t)
&=& \frac{\hs^2}{2\hm}\int_0^1dr\frac{\sqrt{rt}\ e^{\frac{rt}{2}\hd^2}}{1-r}I_1(\hm\sqrt{rt})
I_2(\hs\sqrt{t(1-r)}) \nn\\
&&+e^{\frac{t}{2} \hd^2}\left[
I_0(\hm\sqrt{t})+\frac{\sqrt{t}I_1(\hm\sqrt{t})}{\hm}
\Big(\hd^2+\frac{\hs^2}{2}\Big)\right],\nn\\
Q_S^{\nu=2}(\hm;t)&=&\frac{\hs^3}{2\hm^2}\int_0^1dr\frac{r\sqrt{t}\ e^{\frac{rt}{2}\hd^2}}{(1-r)^{\frac32}}
I_2(\hm\sqrt{rt}) I_3(\hs\sqrt{t(1-r)})\nn\\
&&+ e^{\frac{t}{2} \hd^2}\left[I_0(\hm\sqrt{t})
+\frac{\sqrt{t}}{\hm}I_1(\hm\sqrt{t})\Big(\frac{\hs^2}{2}+2\hd^2\Big)
+\frac{t}{\hm^2}I_2(\hm\sqrt{t})\Big(\frac{\hs^4}{8}+\frac{\hs^2\hd^2}{2}+\hd^4\Big)\right]\ .
\nn
\eea
The fact that we get
the following relation $\lim_{\mu_{1,2}\to0}(-)^nQ_n^\nu(M_2^2)/n!=L_n^0(-N(m_2^2+s^2))$ from eq. (\ref{Qdef}) in the limit of vanishing chemical potentials, implies a
series of generalised Sonine identities, e.g. for $\nu=1$:
\be
\frac{s^2}{2m}\left( \int_0^1 dr\frac{\sqrt{r}}{1-r}I_1(m\sqrt{r})I_2(s\sqrt{1-r})\
+\ I_1(m)\right)\ +\ I_0(m)=I_0(\sqrt{m^2+s^2})\ ,
\ee
etc. The general identity obtained in this way is stated
in the following appendix and proven by induction.

The asymptotic limit of $Q_S^\nu$ then easily translates into the limiting form of the
microscopic kernel:
\be
\lim_{N\to\infty}\frac1N{\cal K}_{N}^\nu(M_1^2,M_2^2)\equiv
{\cal K}_S^\nu(\hm_1,\hm_2)\ =\
\frac12 \int_0^1dt I_0(\hm_1\sqrt{t})Q_S^\nu(\hm_2;t)\ .
\label{Koldnu}
\ee

\begin{figure*}[h]
  \unitlength1.0cm
  \epsfig{file=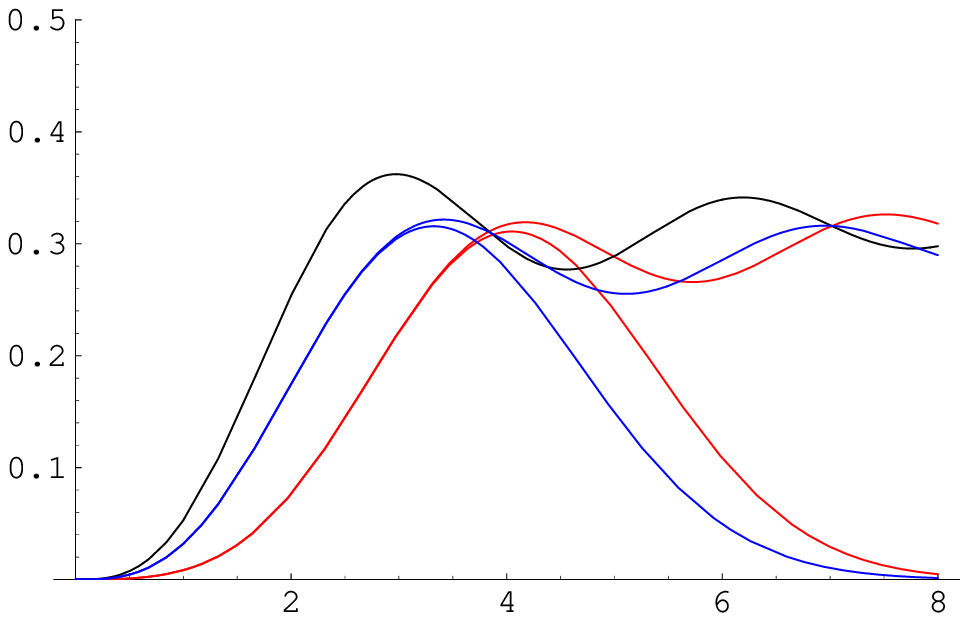,clip=,width=8cm}
    \epsfig{file=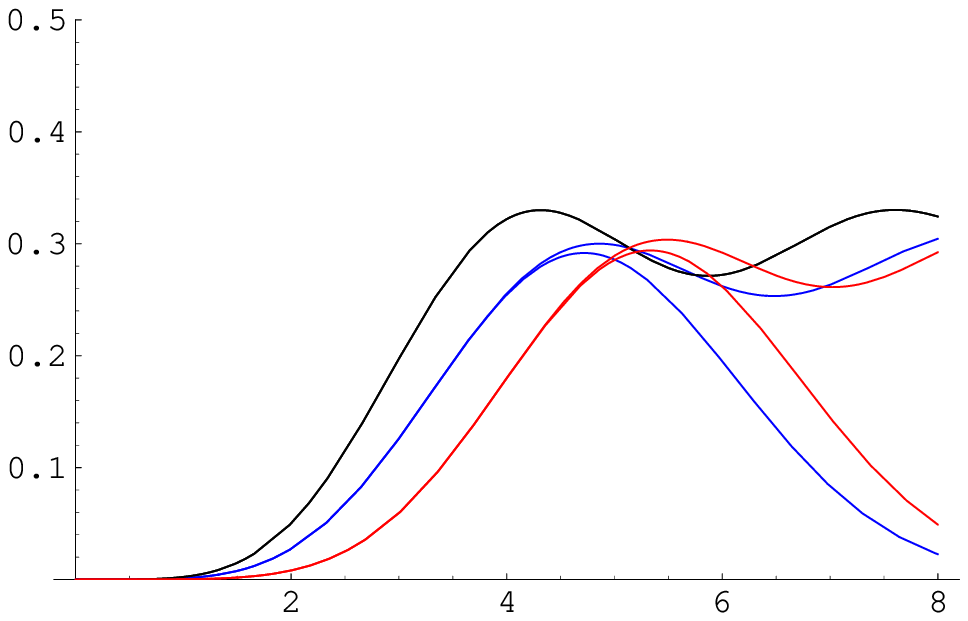,clip=,width=8cm}
  \caption{
    \label{fig:pqnf1nu1-2}
The partially quenched case with $N_1=0$ and $N_2=1$, at topology $\nu=1$ (left plot)
and $\nu=2$ (right plot).
The first eigenvalue is compared with the density eq. (\ref{rho2MM01}), for $\hm=1.5$ at $\hd=2.5$ (middle blue curves)
and for $\hm=1.5$ at $\hd=0.5$ (right red curves). The latter is almost indistinguishable from the $N_f=1$ flavour one-matrix
model result \cite{DNW,DN} (not shown), to which the two-matrix model reduces at $\hd\to0$. We
also display the quenched one-matrix model density
from eq. (\ref{rhoQ}) at the corresponding
value of $\nu=1$ and 2 for comparison (left black curves).
This curve is approached both when $\hm\gg1$ or $\hd\gg1$.
}
\end{figure*}

We are now ready to give some explicit examples for gap probabilities at non-zero topology.
The first example is for one flavour, the simplest partially quenched case, with $N_1=0$ and $N_2=1$ with mass $\hm$. While for $\nu=0$ we had \cite{AD08}
\be
E_{S,\nu=0 }^{(0+1)}(\hs)\ =\ \exp\left[-\frac14\hs^2- \frac12\hd^2\right]
\frac{Q_S^{\nu=0}(\hm;t=1)}{I_0(\hm)} \ ,
\label{E2MM01}
\ee
our new results for topology $\nu=1,2$ are
\bea
E_{S,\nu=1}^{(0+1)}(\hs)&=& \exp\left[-\frac14\hs^2- \frac12\hd^2\right]
\frac{\hm}{I_1(\hm)} {\cal K}_S^{\nu=1}(\hs,\hm)\ ,\\
E_{S,\nu=2}^{(0+1)}(\hs)&=& \exp\left[-\frac14\hs^2- \frac12\hd^2\right]
\frac{\hm^2}{\hs I_2(\hm)}
\left|\begin{array}{ll}
{\cal K}_S^{\nu=2}(\hs,\hm)&\partial_{p}{\cal K}_S^{\nu=2}(p,\hm)|_{p=\hs}\\
I_0(\hs) & I_1(\hs)
\end{array}\right|.
\eea
Note that the normalising partition function is chosen such that it
does not vanish at zero mass, e.g.
${\cal Z}_\nu^{(0+1)}(m)=I_\nu(m)/m^\nu$. It also gives rise to an extra factor $e^{\hd^2}$ per unpaired flavour.
The corresponding first eigenvalues are shown in
figure \ref{fig:pqnf1nu1-2} together with the corresponding densities
from \cite{ADOS}:
\be
\rho_\nu^{(0+1)}(\hx)\ =\ \rho_\nu^{quen}(\hx)\ -\
\exp\left[-\frac12\hd^2\right]\hx
\frac{ J_\nu(\hx)}{I_\nu(\hm_1)} \int_0^1dTT e^{\frac12 T^2 \hd^2}I_\nu(T\hm_1)
J_\nu(T\hx)\ ,
\label{rho2MM01}
\ee
where the quenched one-matrix model spectral density reads \cite{Jacetal}
\be
\rho_\nu^{quen}(\hx)\equiv\frac{\hx}{2}\left(J_\nu(\hx)^2-J_{\nu-1}(\hx)J_{\nu+1}(\hx)\right)\ .
\label{rhoQ}
\ee
It can be seen that our new results for individual eigenvalue distributions nicely follow the corresponding known spectral densities almost up to the first local maximum, for the two different values of topology and various values of $\hm$ and $\hd$ chosen.

\begin{figure*}[h]
  \unitlength1.0cm
  \epsfig{file=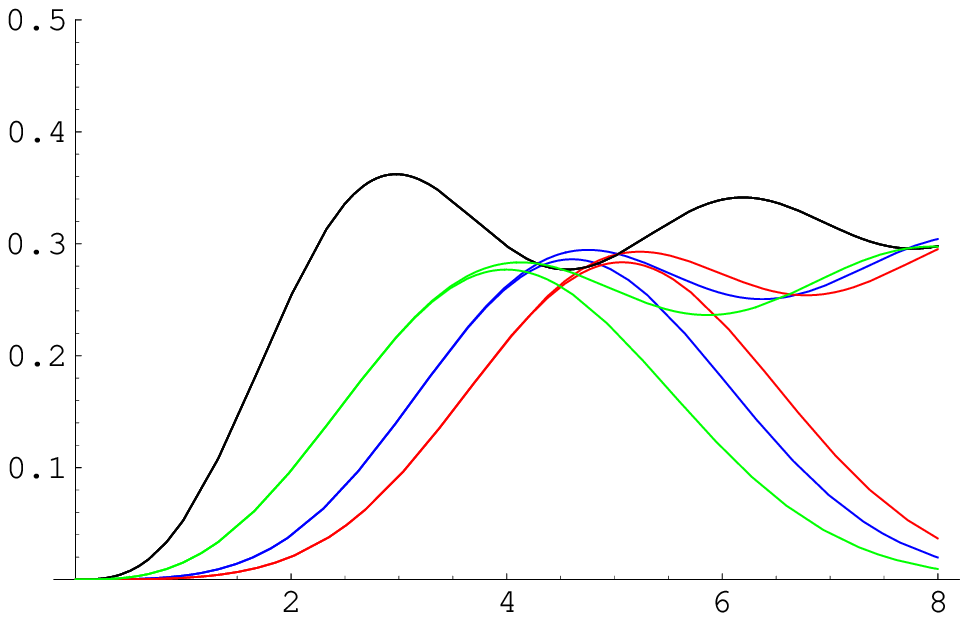,clip=,width=8cm}
    \epsfig{file=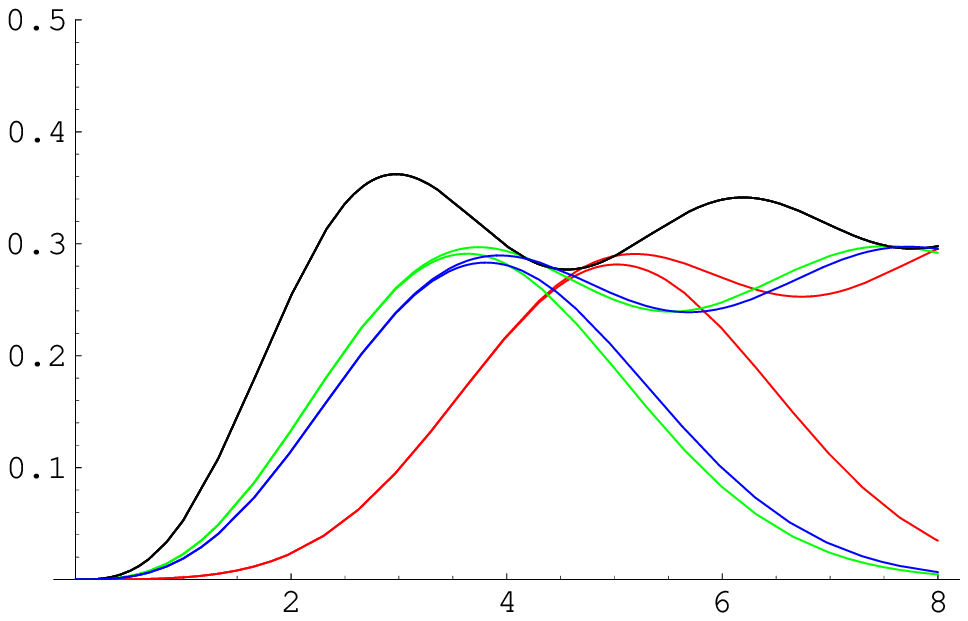,clip=,width=8cm}
  \caption{
    \label{fig:nf2nu1}
The example of two flavours: $N_1=N_2=1$ (left plot), and the
partially quenched case with $N_1=0$ and $N_2=2$ (right plot), both at
topology $\nu=1$.
The first eigenvalue is compared with the corresponding two-matrix model densities, for the same values of parameters:
$\hm_1=1.5$, $\hm_2=3.5$
at $\hd=0.5$ (right red curves), the same masses at $\hd=2.5$ (middle
blue curves),
and $\hm_1=5.5$, $\hm_2=3.5$ at $\hd=2.5$ (left green curves). The
quenched one-matrix model density
from eq. (\ref{rhoQ}) at $\nu=1$ (top black curves) is shown to guide the eye.
The curves at smallest $\hd=0.5$ (right red curves) almost agree for
both flavour settings, being very close to the
the $N_f=2$ flavour one-matrix model result \cite{DNW} (not shown). However, the
other two curves differentiate between $N_f=1+1$ and $N_f=0+2$ flavours
for the same parameter values.
}
\end{figure*}

The second set of examples is the gap for two flavours. First we give the case $N_1=1=N_2$ at
topology $\nu=1$ (and 0):
\bea
E_{S,\nu=0}^{(1+1)}(\hs)&=&\frac{e^{-\hs^2/4}}{{\cal Z}_{\nu=0}^{(1+1)}(\hm_1;\hm_2)}
{\cal K}_S^{\nu=0}(\hm_1^{\prime},\hm_2)\ ,\nn\\
E_{S,\nu=1}^{(1+1)}(\hs)&=&
\frac{e^{-\hs^2/4}}{{\cal Z}_{\nu=1}^{(1+1)}(\hm_1;\hm_2)(\hm_1^{\prime\,2}-s^2)}
\left|\begin{array}{cc}
{\cal K}_S^{\nu=1}(\hm_1^{\prime},\hm_2) &{\cal K}_S^{\nu=1}(s,\hm_2)\\
I_0(\hm_1') & I_0(s)\\
\end{array}\right|,\\
\mbox{with}&&{\cal Z}_{\nu}^{(1+1)}(\hm_1;\hm_2)
=\frac{1}{2(\hm_1\hm_2)^\nu}\int_0^1dt\ e^{\hd^2 t/2}I_\nu(\hm_1\sqrt{t})I_\nu(\hm_2\sqrt{t})\ .
\eea
The corresponding first eigenvalue distributions obtained by differentiating these quantities with respect to $s$ are plotted in figure
\ref{fig:nf2nu1} against the corresponding spectral density \cite{ADOS},
\bea
\rho_\nu^{(1+1)}(\hx)&=& \rho_\nu^{quen}(\hx)-\hx \
\frac{\int_0^1dttJ_\nu(t\hx)I_\nu(t\hm_1)\int_0^1dtt\ e^{\frac12 t^2  \hd^2}
J_\nu(t\hx)I_\nu(t\hm_2)}{
\int_0^1dtt e^{\frac12 t^2  \hd^2}I_\nu(t\hm_1)I_\nu(t\hm_2)}\ .
\label{rho2MMN11}
\eea
Here the difference between $1+1$ flavours (left) and partial quenching (right) becomes visible. Once more the individual eigenvalue distributions agree nicely with the corresponding spectral densities.

The partially quenched two-flavour case, with $N_1=0$ and $N_2=2$ reads as follows in the large-$N$ limit:
\be
E_{\nu=1}^{(0+2)}(s) =\frac{e^{-\hs^2/4-\hd^2}\hm_1\hm_2}{\hm_1I_2(\hm_1)I_1(\hm_2)-\hm_2I_2(\hm_2)I_1(\hm_1)}
\left|
\begin{array}{ll}
{\cal K}_{S}^{\nu=1} (\hs,\hm_1) &{\cal K}_{S}^{\nu=1} (\hs,\hm_2)\\
Q_{S}^{\nu=1}(\hm_1;t=1) & Q_{S}^{\nu=1}(\hm_2;t=1)\\
\end{array}
\right|.
\ee
Note again the proper normalisation of the partition function in masses and $\hd$. The corresponding density to
compare with was derived in \cite{ADOS} where it is given as a
determinant (including the corrections in the Erratum)
\bea
\rho_\nu^{(N_f=0+2)}(\hx)&=& \rho^{quen}_\nu(\hx)
-\ \exp\left[-\frac12\hd^2\right] \hx\
\Big(\hm_1 I_{\nu+1}(\hm_1)I_\nu(\hm_2)-\hm_2I_\nu(\hm_1)I_{\nu+1}(\hm_2)\Big)^{-1}
\label{rho2MM02}\\
&\times&\left[
 \int_0^1dt t e^{\frac12\hd^2 t^2}J_\nu(\hx t)I_\nu(\hm_1 t)
\Big(-I_\nu(\hm_2)(\hx J_{\nu+1}(\hx)+\hd^2J_\nu(\hx)) -
\hm_2 I_{\nu+1}(\hm_2)J_\nu(\hx)\Big)
\right.
\nn\\
&&\left.+
\int_0^1dt t e^{\frac12\hd^2 t^2}J_\nu(\hx t)I_\nu(\hm_2 t)
\Big(I_\nu(\hm_1)(\hx J_{\nu+1}(\hx)+\hd^2J_\nu(\hx))+
\hm_1I_{\nu+1}(\hm_1)J_\nu(\hx)\Big)\right],
\nn
\eea
In all cases where we could compare to the approach presented in the
main body of this paper the curves obtained from the two equivalent
approaches agree.

\sect{Sonine Identities}\label{Sonine}

In this appendix we prove some integral identities relevant for the appendix
\ref{oldnugen} that have to be satisfied and that
can be reduced to the so-called Sonine integral eq. (\ref{SonineId}) below.

We mentioned already that as a check
in the limit
of vanishing chemical potential
the new polynomials $Q_n^\nu$ have to reduce to the
Laguerre polynomials at shifted mass, in order to reproduce the known one-matrix models results.
While for finite-$N$ this can be done
using the first line of the definition eq. (\ref{Qdef}), this is not so easy
after taking the large-$N$ limit. In fact from eq. (\ref{Qasymp}) at $\hd=0$ and $t=1$
\footnote{The case $t\neq1$ is easily reestablished when rescaling
$\hm,\hs\to\sqrt{t}\hm,\sqrt{t}\hs$.} we get the following identities for $\nu=0,1,2$:
\bea
\label{SonineId}
{\cal I}_0&\equiv&
s\int_0^1 dx\frac{x}{\sqrt{1-x^2}}I_0(mx)I_1(s\sqrt{1-x^2}) + I_0(m)-I_0(\sqrt{m^2+s^2})=0\ ,\\
{\cal I}_1&\equiv&
\frac{s^2}{m}\int_0^1 dx\frac{x^2}{1-x^2}I_1(mx)I_2(s\sqrt{1-x^2})
+\frac{s^2}{2m}I_1(m)+ I_0(m)-I_0(\sqrt{m^2+s^2})=0\ ,\\
{\cal I}_2&\equiv&\frac{s^3}{m^2}\int_0^1 dx\frac{x^3}{(1-x^2)^{\frac32}}I_2(mx)I_3(s\sqrt{1-x^2})
+\frac{s^4}{8m^2}I_2(m)+\frac{s^2}{2m}I_1(m)+ I_0(m)-I_0(\sqrt{m^2+s^2})\nn\\
&=&0\ .
\eea
The first equation is the known Sonine identity, see e.g. \cite{Prudnikov}, \cite{Watson} as well as \cite{AD08} for an independent derivation.
It is easier to state and prove the difference between two consecutive integral identities
of this kind:
\bea
0\ =\ {\cal I}_{\nu}-{\cal I}_{\nu-1}&=&\frac{s^{\nu+1}}{m^\nu}\int_0^1 dx\frac{x^{\nu+1}}{(1-x^2)^{\frac{\nu+1}{2}}}
I_{\nu}(mx)I_{\nu+1}(s\sqrt{1-x^2})
+\frac{s^{2\nu}}{2^\nu \nu!\,m^\nu}I_\nu(m)-\nn\\
&&-\ \frac{s^{\nu}}{m^{\nu-1}}\int_0^1 dx\frac{x^{\nu}}{(1-x^2)^{\frac{\nu}{2}}}
I_{\nu-1}(mx)I_{\nu}(s\sqrt{1-x^2})\ .
\label{recnu}
\eea
Because the way to prove the induction start ${\cal I}_{1}-{\cal I}_0=0$ and the induction step
is the same, using integration by parts, we will be brief. If we use
\be
\partial_y\left( \frac{I_{\nu}(y)}{y^\nu}\right)=
\frac{I_{\nu+1}(y)}{y^\nu}\ ,
\ee
upon choosing $y=s\sqrt{1-x^2}$ we can write
\be
\frac{xI_{\nu+1}(s\sqrt{1-x^2})}{(1-x^2)^{(\nu+1)/2}}\ =
\ -\partial_x \left(\frac{I_{\nu}(s\sqrt{1-x^2})}{s(1-x^2)^{\nu/2}}\right).
\ee
We can therefore rewrite the first integral in eq. (\ref{recnu}) as follows:
\bea
&&\int_0^1 dx\frac{x^{\nu+1}}{(1-x^2)^{\frac{\nu+1}{2}}}
I_{\nu}(mx)I_{\nu+1}(s\sqrt{1-x^2})\ =\ \nn\\
&=&\left.-\ \frac{x^\nu I_{\nu}(mx)I_{\nu}(s\sqrt{1-x^2})}{s(1-x^2)^{\nu/2}}\right|_{0}^1
+\int_0^1 dx\frac{I_{\nu}(s\sqrt{1-x^2})}{s(1-x^2)^{\nu/2}}x^\nu m I_{\nu-1}(mx)\nn\\
&=& -\frac{I_\nu(m)}{s}\left(\frac{s}{2}\right)^\nu\frac{1}{\nu!} +
\frac{m}{s}\int_0^1 dx\frac{x^{\nu}}{(1-x^2)^{\frac{\nu}{2}}}
I_{\nu-1}(mx)I_{\nu}(s\sqrt{1-x^2})\ ,
\eea
where we have used a Bessel identity $pI_\nu(p)^\prime+\nu I_\nu(p)=pI_{\nu-1}(p)$ as well as
the series representation of the Bessel function to determine the limit at the upper bound $x=1$.
Inserting this for $\nu=1$ and then general $\nu$ yields the induction start and
induction step.

\end{appendix}

\end{document}